\documentclass[twocolumn]{aastex631}

\usepackage{hyperref,enumerate}
\usepackage{xcolor}

\hypersetup{
    unicode=false,                  % non-Latin characters in Acrobat book
    pdftoolbar=true,                % show Acrobat toolbar?
    pdfmenubar=true,                % show Acrobat menu?
    pdffitwindow=true,              % window fit to page when opened
    pdfstartview={FitH},            % fits the width of the page t
    pdftitle={SPLUSJ1424-2542},     % title
    pdfauthor={vplacco},            % author
    pdfsubject={Astronomy},         % subject of the document
    pdfcreator={dvipdf},            % creator of the document
    pdfproducer={dvipdf},           % producer of the document
    pdfkeywords={metal-poor stars}, % list of keywords
    pdfnewwindow=true,              % links in new window
    colorlinks=true,                % false: boxed links; true: colored
    linkcolor=red,                  % color of internal links
    citecolor=blue,                 % color of links to bibliography
    filecolor=magenta,              % color of file links
    urlcolor=cyan,                  % color of external links
    breaklinks=true,
    linktocpage
}

\newcommand{\kmsec}{\mbox{km~s$^{\rm -1}$}}
\newcommand{\eps}[1]{\ensuremath{\log\epsilon\,(\mathrm{#1})}}
\newcommand{\vv}{{\tablenotemark{\footnotesize{a}}}}

\newcommand{\abund}[2]{\ensuremath{[\mathrm{#1}/\mathrm{#2}]}}
\newcommand{\cfe}{\abund{C}{Fe}}

\newcommand{\xfe}[1]{\abund{#1}{Fe}}

\newcommand{\metal}{\abund{Fe}{H}}

\newcommand{\teff}{\ensuremath{T_\mathrm{eff}}}
\newcommand{\logg}{\ensuremath{\log\,g}}

\newcommand{\A}[1]{\ensuremath{A(\mathrm{#1})}}

\newcommand{\K}{\ensuremath{\mathrm{K}}}

\newcommand{\colorx}{{\texttt{(J0395-J0410)-(J0660-J0861)}}}
\newcommand{\colory}{{\texttt{(J0395-J0660)-2$\times$(g-i)}}}

\newcommand{\ump}{\object{SPLUS~J2104$-$0049}}

\newcommand{\rave}{\object{SPLUS~J1424$-$2542}}
\newcommand{\ravel}{\object{SPLUS~J142445.34$-$254247.1}}

\newcommand{\splus}{S-PLUS}

\received{September 22,2023}
\revised{October 24, 2023}
\accepted{October 25,2023}

\submitjournal{ApJ}

\begin{document}

\title{\ravel: An R-Process Enhanced, Actinide-Boost, \\
Extremely Metal-Poor star observed with GHOST
%\footnote{
%
%Based on observations obtained at the International Gemini Observatory, a
%program of NSF’s NOIRLab, which is managed by the Association of Universities
%for Research in Astronomy (AURA) under a cooperative agreement with the
%National Science Foundation on behalf of the Gemini Observatory
%partnership: the National Science Foundation (United States), National
%Research Council (Canada), Agencia Nacional de Investigaci\'{o}n y
%Desarrollo (Chile), Ministerio de Ciencia, Tecnolog\'{i}a e Innovaci\'{o}n
%(Argentina), Minist\'{e}rio da Ci\^{e}ncia, Tecnologia, Inova\c{c}\~{o}es e
%Comunica\c{c}\~{o}es (Brazil), and Korea Astronomy and Space Science
%Institute (Republic of Korea).}  
}

%\suppressAffiliations

\author[0000-0003-4479-1265]{Vinicius M.\ Placco}
\affiliation{NSF’s NOIRLab, Tucson, AZ 85719, USA}

\author[0000-0002-8048-8717]{Felipe Almeida-Fernandes}
\affiliation{Departamento de Astronomia, Instituto de Astronomia, Geof\'isica e
Ci\^encias Atmosf\'ericas da USP, Cidade \\ Universit\'aria, 05508-900, S\~ao
Paulo, SP, Brazil}

\author[0000-0002-5463-6800]{Erika M.\ Holmbeck}
\altaffiliation{NHFP Hubble Fellow}
\affiliation{Observatories of the Carnegie Institution 
for Science, Pasadena, CA 91101, USA}

\author[0000-0001-5107-8930]{Ian U.\ Roederer}
\affiliation{%
Department of Physics, North Carolina State University,
Raleigh, NC 27695, USA}
\affiliation{%
Department of Astronomy, University of Michigan,
Ann Arbor, MI 48109, USA}
\affiliation{%
Joint Institute for Nuclear Astrophysics -- Center for the
Evolution of the Elements (JINA-CEE), USA}

\author[0000-0001-9178-3992]{Mohammad K.\ Mardini}
\affiliation{Department of Physics, Zarqa University, Zarqa 13110, Jordan}
\affiliation{Kavli IPMU (WPI), UTIAS, The University of Tokyo, Kashiwa, Chiba 277-8583, Japan}
\affiliation{Institute for AI and Beyond, The University of Tokyo 7-3-1 Hongo, Bunkyo-ku, Tokyo 113-8655, Japan}

\author[0000-0003-2969-2445]{Christian R.\ Hayes}
\affiliation{NRC Herzberg Astronomy and Astrophysics Research Centre, 5071 West Saanich Road, Victoria, B.C., V9E 2E7, Canada}

\author[0000-0003-4134-2042]{Kim Venn}
\affiliation{Department of Physics and Astronomy, University of Victoria, Victoria, BC, V8W 3P2, Canada}

%%%%%%%%%%%%%%%%%

\author[0000-0002-9020-5004]{Kristin Chiboucas}
\affiliation{Gemini Observatory/NSF’s NOIRLab, 670 N. A’ohoku Place, Hilo, HI, 96720, USA}

\author[0000-0001-9796-2158]{Emily Deibert}
\affiliation{Gemini Observatory/NSF’s NOIRLab, Casilla 603, La Serena, Chile}

\author[0000-0002-5227-9627]{Roberto Gamen}
\affiliation{Instituto de Astrof\'isica de La Plata, CONICET--UNLP, Paseo del Bosque s/n, 1900, La Plata, Argentina}

\author[0000-0003-2530-3000]{Jeong-Eun Heo}
\affiliation{Gemini Observatory/NSF’s NOIRLab, Casilla 603, La Serena, Chile}

\author[0009-0009-7838-7771]{Miji Jeong}
\affiliation{Department of Astronomy, Space Science, and Geology, Chungnam National University, Daejeon 34134, Republic of Korea}

\author{Venu Kalari}
\affiliation{Gemini Observatory/NSF’s NOIRLab, Casilla 603, La Serena, Chile}

\author[0000-0002-5084-168X]{Eder Martioli}
\affiliation{Laborat\'orio Nacional de Astrof\'isica, Rua Estados Unidos 154, 37504-364, Itajub\'a, MG, Brazil}

\author[0000-0002-8808-4282]{Siyi Xu}
\affiliation{Gemini Observatory/NSF’s NOIRLab, 670 N. A’ohoku Place, Hilo, HI, 96720, USA}

%%%%%%%%%%%%%%%%%%%%%%%%%%%%%%%%%%%%%%

\author{Ruben Diaz}
\affiliation{Gemini Observatory/NSF’s NOIRLab, Casilla 603, La Serena, Chile}

\author{Manuel Gomez-Jimenez}
\affiliation{Gemini Observatory/NSF’s NOIRLab, Casilla 603, La Serena, Chile}

\author{David Henderson}
\affiliation{Gemini Observatory/NSF’s NOIRLab, 670 N. A’ohoku Place, Hilo, HI, 96720, USA}

\author{Pablo Prado}
\affiliation{Gemini Observatory/NSF’s NOIRLab, Casilla 603, La Serena, Chile}

\author[0000-0001-5558-6297]{Carlos Quiroz}
\affiliation{Gemini Observatory/NSF’s NOIRLab, Casilla 603, La Serena, Chile}

\author[0000-0001-7518-1393]{Roque Ruiz-Carmona}
\affiliation{Gemini Observatory/NSF’s NOIRLab, Casilla 603, La Serena, Chile}

\author[0000-0001-8589-4055]{Chris Simpson}
\affiliation{Gemini Observatory/NSF’s NOIRLab, 670 N. A’ohoku Place, Hilo, HI, 96720, USA}

\author{Cristian Urrutia}
\affiliation{Gemini Observatory/NSF’s NOIRLab, Casilla 603, La Serena, Chile}

%%%%%%%%%%%%%%%%%%%%%%%%%%%%%%%%%%%%%%

\author[0000-0003-4666-6564]{Alan W. McConnachie}
\affiliation{NRC Herzberg Astronomy and Astrophysics Research Centre, 5071 West Saanich Road, Victoria, B.C., V9E 2E7, Canada}

\author{John Pazder}
\affiliation{NRC Herzberg Astronomy and Astrophysics Research Centre, 5071 West Saanich Road, Victoria, B.C., V9E 2E7, Canada}
\affiliation{Department of Physics and Astronomy, University of Victoria, Victoria, BC, V8W 3P2, Canada}

\author{Gregory Burley}
\affiliation{NRC Herzberg Astronomy and Astrophysics Research Centre, 5071 West Saanich Road, Victoria, B.C., V9E 2E7, Canada}

\author[0000-0002-6194-043X]{Michael Ireland}
\affiliation{Research School of Astronomy and Astrophysics, Australian National University, Canberra 2611, Australia}

\author{Fletcher Waller}
\affiliation{Department of Physics and Astronomy, University of Victoria, Victoria, BC, V8W 3P2, Canada}

\author[0000-0002-2606-5078]{Trystyn A. M. Berg}
\affiliation{Dipartimento di Fisica G. Occhialini, Università degli Studi di Milano Bicocca, Piazza della Scienza 3, I-20126 Milano, Italy}
\affiliation{NRC Herzberg Astronomy and Astrophysics Research Centre, 5071 West Saanich Road, Victoria, B.C., V9E 2E7, Canada}

\author[0000-0001-5528-7801]{J. Gordon Robertson}
\affiliation{Australian Astronomical Optics, Macquarie University, 105 Delhi Rd, North Ryde NSW 2113, Australia}
\affiliation{School of Physics, University of Sydney, NSW 2006, Australia}

%%%%%%%%%%%%%%%%%%%%%%%%%%%%%%%%%%%%%%%

\author[0000-0003-4236-6927]{Zachary Hartman}
\affiliation{Gemini Observatory/NSF’s NOIRLab, 670 N. A’ohoku Place, Hilo, HI, 96720, USA}

\author[0000-0002-6230-0151]{David O. Jones}
\affiliation{Gemini Observatory/NSF’s NOIRLab, 670 N. A’ohoku Place, Hilo, HI, 96720, USA}
\affiliation{Institute for Astronomy, University of Hawai‘i, 640 N.\ Aohoku Pl., Hilo, HI 96720, USA}

\author[0000-0002-6633-7891]{Kathleen Labrie}
\affiliation{Gemini Observatory/NSF’s NOIRLab, 670 N. A’ohoku Place, Hilo, HI, 96720, USA}

\author{Gabriel Perez}
\affiliation{Gemini Observatory/NSF’s NOIRLab, Casilla 603, La Serena, Chile}

\author{Susan Ridgway}
\affiliation{NSF’s NOIRLab, Tucson, AZ 85719, USA}

\author[0000-0003-1033-4402]{Joanna Thomas-Osip}
\affiliation{Gemini Observatory/NSF’s NOIRLab, Casilla 603, La Serena, Chile}

%%%%%%%%%%%%%%%%%%%%%%%%%%%%%%%%%%%%%%%

\correspondingauthor{Vinicius M.\ Placco}
\email{vinicius.placco@noirlab.edu}

\begin{abstract}

We report on the chemo-dynamical analysis of \ravel, an extremely metal-poor halo star enhanced in elements formed by the rapid neutron-capture process. This star was first selected as a metal-poor candidate from its narrow-band S-PLUS photometry and followed up spectroscopically in medium-resolution with Gemini South/GMOS, which confirmed its low-metallicity status. High-resolution spectroscopy was gathered with GHOST at Gemini South, allowing for the determination of chemical abundances for 36 elements, from carbon to thorium. 
At \metal=$-3.39$, \rave\, is one of the lowest metallicity stars with measured Th and has the highest \eps{Th/Eu} observed to date, making it part of the ``actinide-boost'' category of $r$-process enhanced stars.
The analysis presented here suggests that the gas cloud from which \rave\, was formed must have been enriched by at least two progenitor populations. The light-element ($Z\leq30$) abundance pattern is consistent with the yields from a supernova explosion of metal-free stars with 11.3--13.4\,M$_\odot$, and the heavy-element ($Z\geq38$) abundance pattern can be reproduced by the yields from a neutron star merger (1.66\,M$_\odot$ and 1.27\,M$_\odot$) event. A kinematical analysis also reveals that \rave\, is a low-mass, old halo star with a likely in-situ origin, not associated with any known early merger events in the Milky Way.

\end{abstract}


\section{Introduction} 
\label{intro}

%%%%%%%%%%%%%%%%% r-process and things...

The element europium (Eu; $Z=63$), formed mainly by the rapid neutron-capture process \citep[$r$-process;][]{b2fh}, has been identified in the spectrum of the Sun by \citet{dyson1906}, from observations taken during the 1900, 1901, and 1905 total Solar eclipses. In other stars, some of the first measurements of Eu also date back to the early 1900's \citep{lunt1907,baxandall1913}. In fact, \citet{lunt1907} describes europium as a ``disturbing element'' when trying to determine the radial velocities for the $\alpha$-Bo\"otis and $\beta$-Geminorum stars from a calcium absorption feature\footnote{For the purpose of the present work, this calcium line is the actual ``disturbing element'' when performing the spectral synthesis of the Eu $\lambda$4435 line, which is shown in later sections.}. Since then, Eu has established itself as a crucial tracer of the operation of the $r$-process in the Galaxy and beyond, with a large number of measurable absorption features in the optical wavelength regime.

In this context, low-mass, long-lived, old stars in the Galactic halo hold in their atmospheres valuable insights into the nucleosynthesis in the early Universe and the formation of heavy elements. They are the key to understanding the chemical evolution of the Universe. From a theoretical perspective, the nucleosynthesis pathways from hydrogen to the heavy elements (loosely defined as $Z>30$) have been understood for almost 80 years \citep[e.g.][]{hoyle1946}. These heavy elements have also been identified in stellar atmospheres even before \citep[][and references in the paragraph above]{merrill1926}, but it was only in the past 50 years or so that high-resolution spectroscopy was able to quantify the chemical abundances in a statistically relevant and consistent way \citep[][to name a few]{cowley1973,spite1978,luck1981,truran1981,sneden1985,gilroy1988,sneden1994}.
The past 25 years have seen a tremendous increase in the number of high-resolution spectroscopic observations of metal-poor stars (\metal\footnote{\abund{A}{B} = $\log(N_A/{}N_B)_{\star} - \log(N_A/{}N_B) _{\odot}$, where $N$ is the number density of atoms of a given element in the star ($\star$) and the Sun ($\odot$), respectively.} $\lesssim -1.0$) with enhancement in elements formed by the $r$-process, in particular the so-called $r$-II stars \citep[\xfe{Eu}$>+1$\footnote{More recently, \citet{holmbeck2020} has empirically re-defined the $r$-II classification boundary to be \xfe{Eu}$>+0.7$.} and \abund{Ba}{Eu}$<0$;][]{frebel2018}.

The nucleosynthesis of $r$-process elements requires high neutron fluxes and it is believed to occur in extreme astrophysical events, such as the aftermath of neutron star mergers \citep{goriely2011,abbott2017,drout2017,shappee2017} or the evolution of massive stars \citep{siegel2019,grichener2019}, and the subsequent pollution of the interstellar medium by these elements has allowed the formation of such peculiar low-mass $r$-II stars.
Understanding the properties and distribution of such stars is crucial for constraining $r$-process models and gaining insights into the conditions prevalent in the early universe.
Recent studies have also provided insight into the astrophysical environments that would harbor such extreme events and enable the formation of $r$-II stars. As an example, dwarf galaxies and stellar overdensities were found to contain low-metallicity, $r$-process enhanced stars \citep{vincenzo2015,ji2016,hansen2017,roederer2018b,yuan2020,gudin2021,abuchaim2023,shank2023}. 

%%%%%%%%%%%%%%%% selecting metal-poor stars and advertising surveys...

$r$-II stars are not a common occurrence within very metal-poor samples in the Milky Way. The first systematic search for such objects was the Hamburg/ESO R-process Enhanced star Survey \citep[HERES;][]{christlieb2004,barklem2005}, which obtained data for 253 metal-poor halo stars. More recently, the $R$-Process Alliance \citep[RPA;][]{hansen2018,sakari2018,ezzeddine2020,holmbeck2020} has been making outstanding progress in further discovering and analyzing $r$-process enhanced stars. 
Both HERES and RPA adopt a two-step approach, first identifying metal-poor stars from medium-resolution (R$\sim2,000$) spectroscopy \citep{frebel2006,placco2018,placco2019} then collecting ``snapshot'' (S/N$\sim50$ and R$\sim$20,000) spectra for the confirmed candidates. Further studies are then conducted for the most interesting candidates within those samples \citep[][among many others]{jonsell2006,mashonkina2010,ren2012,cui2013,mashonkina2014,mashonkina2014b,hill2017,placco2017,cain2018,gull2018,holmbeck2018,roederer2018,sakari2018b,placco2020,roederer2022}. Even within those somewhat targeted searches, the fraction of $r$-II stars (\xfe{Eu}$>+1.0$) found in HERES is 3\%, while for the RPA is 8\%, using data from their four ``data release'' articles mentioned above. There is a clear need for continuing the identification of such objects in order to properly constrain their occurrence fractions and astrophysical sites.

%%%%%%%%%%%%%%%%% wrapping up

In this article, we continue in the quest to increase the number of identified $ r$-process-enhanced stars in the Milky Way. We present the chemo-dynamical analysis of \ravel\, (hereafter \rave) using data from the recently commissioned GHOST spectrograph at Gemini South. At \metal$=-3.39$ with a low carbon-to-iron abundance ratio, \rave\, has a distinctive $r$-process signature with an enhancement in thorium when compared to the scaled Solar System $r$-process abundance pattern. From a kinematics perspective, \rave\, is a low-mass, old halo star with a probable {\emph{in-situ}} origin.

This work is outlined as follows: Section~\ref{selection} details the target selection and observations, followed by the determination of stellar atmospheric parameters and chemical abundances in Section~\ref{atmpars}. In Section~\ref{chemod} we analyze the chemical abundance pattern of \rave\, and its dynamical properties, aiming to infer characteristics of the progenitor population(s). Final remarks and perspectives for future work are presented in Section~\ref{conclusion}.

\section{Target selection and Observations}
\label{selection}

In this section, we briefly describe the identification, selection, and spectroscopic follow-up observations of \rave. Table~\ref{starinfo} lists basic information and derived quantities for \rave, measured in this work and other studies in the literature\footnote{\rave\, has been independently observed as part of the SkyMapper Southern Survey Data Release 1 \citep[SMSS DR1;][]{wolf2018} as SMSS~J142445.33$-$254246.9. It was followed up with medium-resolution spectroscopy (R$\sim$3,000) as part of the search for extremely metal-poor stars conducted by \citet{dacosta2019}. For reference, the atmospheric parameters determined by \citet{dacosta2019} are provided in Table~\ref{starinfo}.}. Further details can also be found in \citet{Placco+2022}.

%https://simbad.u-strasbg.fr/simbad/sim-id?Ident=%4021454261&Name=SMSS+J142445.33-254246.9

\begin{deluxetable*}{lllll}
\tablecaption{Properties of \protect\ravel \label{starinfo}}
\tablewidth{0pt}
\tabletypesize{\scriptsize}
\tabletypesize{\small}
\tablehead{
\colhead{Quantity} &
\colhead{Symbol} &
\colhead{Value} &
\colhead{Units} &
\colhead{Reference}}
\startdata
Right ascension            & $\alpha$ (J2000)    & 14:24:45.34               & hh:mm:ss.ss       & \citet{gaia23dr3}             \\ %OK
Declination                & $\delta$ (J2000)    & $-$25:42:47.1             & dd:mm:ss.s        & \citet{gaia23dr3}             \\ %OK
Galactic longitude         & $\ell$              & 327.983                   & degrees           & \citet{gaia23dr3}             \\ %OK
Galactic latitude          & $b$                 & 32.579                    & degrees           & \citet{gaia23dr3}             \\ %OK
Gaia DR3 ID                &                     & 6271613367058424064       &                   & \citet{gaia23dr3}             \\ %OK
Parallax                   & $\varpi$            & 0.0796 $\pm$ 0.0182       & mas               & \citet{gaia23dr3}             \\ %OK
Inverse parallax distance  & $1/\varpi$          & 8.13$^{+1.41}_{-1.05}$    & kpc               & This study\vv                 \\ %OK
Distance                   & $d$                 & 7.82$^{+0.95}_{-0.76}$    & kpc               & \citet{Bailer-Jones+2021}     \\ %OK
Proper motion ($\alpha$)   & PMRA                & $-$2.643 $\pm$ 0.022      & mas yr$^{-1}$     & \citet{gaia23dr3}             \\ %OK
Proper motion ($\delta$)   & PMDec               &    0.956 $\pm$ 0.026      & mas yr$^{-1}$     & \citet{gaia23dr3}             \\ %OK
$K$ magnitude              & $K$                 & 11.538 $\pm$ 0.021        & mag               & \citet{skrutskie2006}         \\ %OK
$G$ magnitude              & $G$                 & 13.794 $\pm$ 0.003        & mag               & \citet{gaia23dr3}             \\ %OK
$BP$ magnitude             & $BP$                & 14.340 $\pm$ 0.003        & mag               & \citet{gaia23dr3}             \\ %OK
$RP$ magnitude             & $RP$                & 13.087 $\pm$ 0.004        & mag               & \citet{gaia23dr3}             \\ %OK
$g$ magnitude              & $gSDSS$             & 14.435 $\pm$ 0.002        & mag               & \citet{almeida-fernandes2022} \\ %OK
\colorx                    &                     & $-$0.155 $\pm$ 0.011      & mag               & \citet{Placco+2022}           \\ %OK
\colory                    &                     & $-$0.315 $\pm$ 0.009      & mag               & \citet{Placco+2022}           \\ %OK 
Color excess               & $E(B-V)$            & 0.0647 $\pm$ 0.0021       & mag               & \citet{schlafly2011}          \\ %OK
Bolometric correction      & BC$_V$              & $-$0.59 $\pm$ 0.09        & mag               & \citet{casagrande2014}        \\ %OK
\hline
Signal-to-noise ratio           @3860\AA  & S/N  & 28                        & pixel$^{-1}$      & This study (GHOST)            \\ %OK
\phantom{Signal to noise ratio} @4360\AA        && 54                        & pixel$^{-1}$      & This study (GHOST)            \\ %OK
\phantom{Signal to noise ratio} @5180\AA        && 139                       & pixel$^{-1}$      & This study (GHOST)            \\ %OK
\phantom{Signal to noise ratio} @6540\AA        && 171                       & pixel$^{-1}$      & This study (GHOST)            \\ %OK
\hline
Effective Temperature      & \teff               & 4700 $\pm$ 150            & K                 & \citet{Placco+2022} (GMOS)    \\ %OK
                           &                     & 4750                      & K                 & \citet{dacosta2019}           \\ %OK
                           &                     & 4762 $\pm$ 36             & K                 & This study (GHOST)            \\ %OK
Log of surface gravity     & \logg               & 1.48 $\pm$ 0.20           & (cgs)             & \citet{Placco+2022} (GMOS)    \\ %OK
                           &                     & 1.00                      & (cgs)             & \citet{dacosta2019}           \\ %OK
                           &                     & 1.58 $\pm$ 0.11           & (cgs)             & This study (GHOST)            \\ %OK
Microturbulent velocity    & $\xi$               & 1.60 $\pm$ 0.20           & \kmsec            & This study (GHOST)            \\ %OK
Metallicity                & \metal              & $-$3.82 $\pm$ 0.20        & dex               & \citet{Placco+2022} (GMOS)    \\ %OK
                           &                     & $-$3.25                   & dex               & \citet{dacosta2019}           \\ %OK
                           &                     & $-$3.39 $\pm$ 0.12        & dex               & This study (GHOST)            \\ %OK
%                          &                     & $-$X.XX $\pm$ 0.XX        & dex               & This study NLTE (GHOST)       \\ %OK
\hline
Age                        &                     & $10.09_{-3.12}^{+2.96}$   & Gyr               & \citet{AlmeidaFernandes+2023} \\ %OK
Mass                       & $M$                 & $0.843_{-0.056}^{+0.079}$ & $M_{\odot}$       & \citet{AlmeidaFernandes+2023} \\ %OK
Radial velocity            & RV                  & $-$31.2 $\pm$ 0.5         & \kmsec            & This study (MJD: 60074.25416667) \\ %OK
Galactocentric coordinates & ($X,Y,Z$)           & ($+2.61,-3.50,+4.22$)     & kpc               & This study                   \\ %OK
Galactic space velocity    & ($U,V,W$)           & ($-93.0,-29.4,+46.4$)     & km s$^{-1}$       & This study                   \\ %OK
Total space velocity       & $V_\mathrm{Tot}$    & $+108.0$                  & km s$^{-1}$       & This study                   \\ %OK
Apogalactic radius         & $R_\mathrm{apo}$    & $+$8.43 $\pm$ 1.08        & kpc               & This study                   \\ %OK
Perigalactic radius        & $R_\mathrm{peri}$   & $+$5.09 $\pm$ 0.51        & kpc               & This study                   \\ %OK
Max. distance from the Galactic plane & $z_\mathrm{max}$  & $+$6.48 $\pm$ 1.90 & kpc             & This study                   \\ %OK
Orbital eccentricity       & $e$                 & $+$0.25 $\pm$ 0.03        &                   & This study                   \\ %OK
Vertical angular momentum  & $L_Z$               & $+0.849 \pm 0.161 \cdot 10^3$  & kpc km s$^{-1}$   & This study              \\ %OK
Total orbital energy       & $E$                 & $-1.640 \pm 0.525 \cdot 10^5$  & km${^2}$ s$^{-2}$ & This study              \\ %OK
\enddata
\tablenotetext{a}{Using $\varpi_{\rm zp} = -0.0434$ mas from \citet{lindegren2020}.} %OK
\end{deluxetable*}

\begin{figure*}
 \includegraphics[width=1\linewidth]{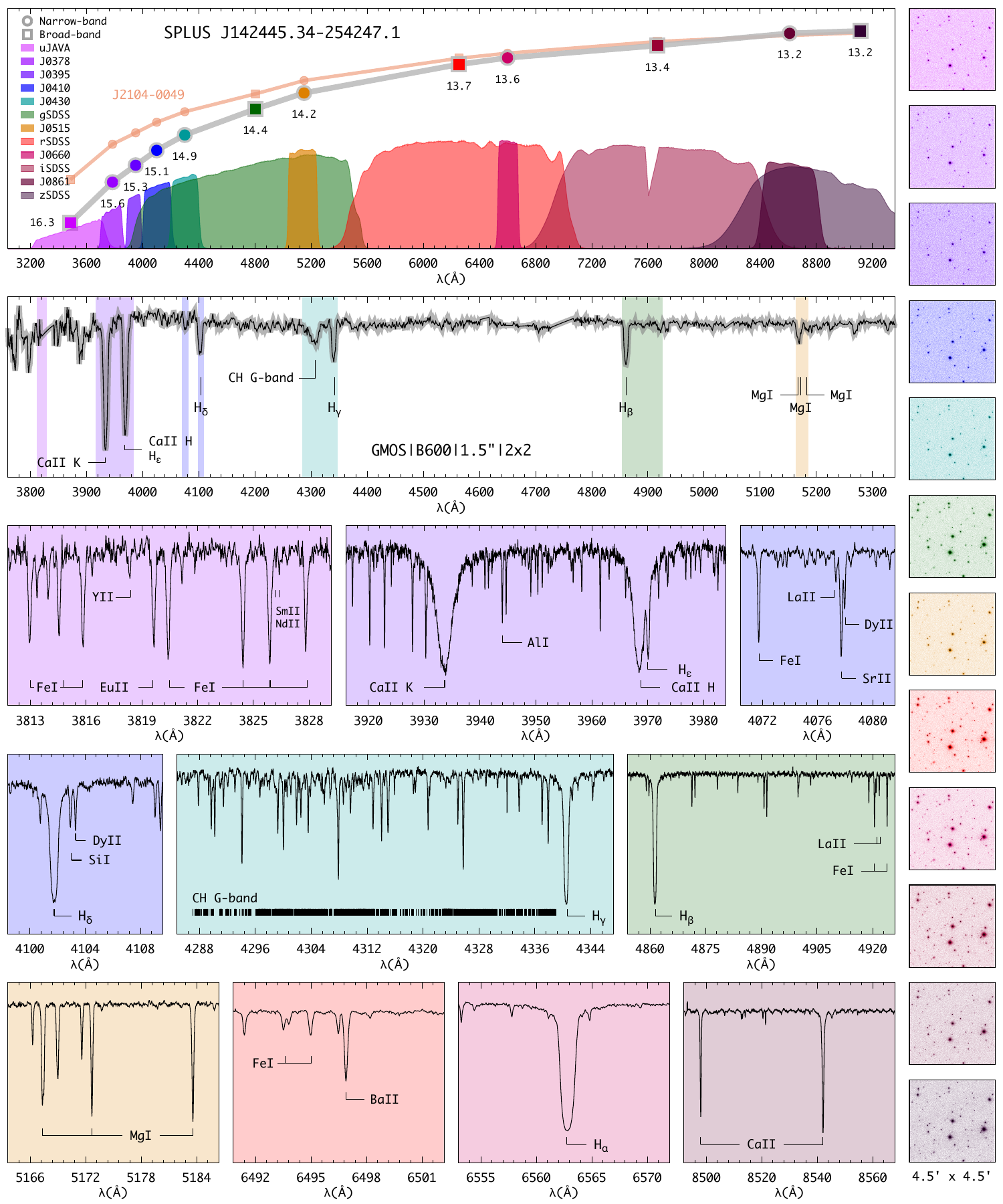}
     \caption{Top: S-PLUS magnitudes for \protect{\rave}\, compared to the values for \ump. Also shown are the S-PLUS filter transmission curves. The second panel from top to bottom shows the Gemini/GMOS spectrum, highlighting absorption features of interest. The remaining color panels show sections of the GHOST spectra and features used for chemical abundance determinations. The 12 side panels show the S-PLUS images for \protect{\rave}. RGB colors in all panels are assigned based on the central wavelength of each filter. See text for further details.}
 \label{combo}
\end{figure*}

\subsection{S-PLUS and Gemini/GMOS}

\rave\, was observed as part of the Southern Photometric Local Universe Survey \citep[\splus;][]{mendesdeoliveira2019} second data release \citep[DR2;][]{almeida-fernandes2022}. \splus\, has a unique 12 broad- and narrow-band filter set, consisting of four SDSS ($g$, $r$, $i$, $z$), one modified SDSS $u$, and seven narrow-band filters. \rave\, was selected as a metal-poor star candidate by \citet{Placco+2022}, based on its narrow-band metallicity-sensitive colors. 
These colors, \colorx\, and \colory, are listed in Table~\ref{starinfo} and place \rave\, in the same regime as other spectroscopically confirmed low-metallicity stars \citep[cf. Figures 1 and 7 of][]{Placco+2022}. In \citet{AlmeidaFernandes+2023}, four criteria for the selection of metal-poor stars from \splus\, were proposed, resulting in different levels of completeness and purity. We note that \rave\, was selected as a low metallicity candidate in all the considered cases.

The top panel of Figure~\ref{combo} shows the \splus\, filter curves, and the twelve magnitudes (AB system) for \rave. Image cutouts for each filter ($4.5'$x$ 4.5'$ centered at \rave) are shown on the right side of the figure. RGB colors are assigned based on the central wavelength of each filter. As a comparison, the \splus\, magnitudes (scaled to the {\texttt{zSDSS}} value for \rave) for \ump, an ultra metal-poor star with \metal=$-4.03$ \citep{placco2021b}, are shown. Both stars have similar temperatures, meaning that the differences in flux for the blue filters can be attributed to lower emerging flux for \rave\, due to the presence of absorption features, a consequence of its overall higher chemical abundances when compared to \ump.

Medium-resolution (R~$\sim1,500$) spectroscopy for \rave\, was gathered on June 18, 2021, with the 8.1\,m Gemini South telescope and the GMOS \citep[Gemini Multi-Object Spectrograph;][]{davies1997,gimeno2016} instrument, as part of the Poor Weather program GS-2021A-Q-419. Further details on the observing setup and data reduction are given in \citet{Placco+2022}. The second panel from top to bottom on Figure~\ref{combo} shows the normalized GMOS data, highlighting a few absorption features of interest for the determination of the effective temperature (\teff\, -- Balmer lines H$\beta$, H$\gamma$, and H$\delta$),  metallicity (\metal --\ion{Ca}{2}~K), carbon abundance (CH G-band), and $\alpha$-element abundance (\ion{Mg}{1}~b triplet). The atmospheric parameters determined by \citet{Placco+2022} are provided in Table~\ref{starinfo}. Based on these parameters, \rave\, was selected as a potential candidate for high-resolution spectroscopic follow-up.

\subsection{Gemini/GHOST}

\rave\, was followed up in high resolution using the newly commissioned GHOST \citep[Gemini High-resolution Optical SpecTrograph;][]{ireland2014,mcconnachie2022,hayes2023} at Gemini South. Observations were conducted on May 10, 2023, as part of the GHOST SV (System Verification\footnote{\href{https://www.gemini.edu/instrumentation/ghost/ghost-system-verification}{https://www.gemini.edu/instrumentation/ghost/ghost-system-verification}} - Program ID: GS-2023A-SV-101) and the data is publicly available at the Gemini Observatory Archive\footnote{\href{https://archive.gemini.edu/searchform/GS-2023A-SV-101-9/}{https://archive.gemini.edu/searchform/GS-2023A-SV-101-9/}}. The instrument setup chosen was the standard resolution (SR: R~$\sim50,000$) and target mode IFU1:Target|IFU2:Sky. For both the blue and red cameras, six 900-second exposures were taken with a 1x2 binning (spectral x spatial). During the observations, the image quality (IQ) and cloud cover (CC) were in the 70$^{th}$-percentile and the sky background (SB) was in the 50$^{th}$-percentile\footnote{Further details on the observing constraints can be found at \href{https://gemini.edu/observing/telescopes-and-sites/sites}{https://gemini.edu/observing/telescopes-and-sites/sites}}.
The wavelength coverage is [3474:5438] \AA\, for the blue camera and [5209:10608] \AA\, for the red camera.

The data reduction was performed using v3.0 of the {\texttt{DRAGONS}}\footnote{\href{https://github.com/GeminiDRSoftware/DRAGONS}{https://github.com/GeminiDRSoftware/DRAGONS}.} software package \citep{dragons,labrie2022}. This version includes support for GHOST, based on the GHOST Data Reduction pipeline v1.0 \citep[\texttt{GHOST DR} - originally described in][]{ireland2018,hayes2022}, which was 
modified by the \texttt{DRAGONS} team during the commissioning of GHOST.
% The data reduction was performed using v3.0 of the {\texttt{DRAGONS}}\footnote{\href{https://github.com/GeminiDRSoftware/DRAGONS}{https://github.com/GeminiDRSoftware/DRAGONS}.} software package \citep{dragons,labrie2022}. This version includes the GHOST Data Reduction pipeline v1.0 \citep[\texttt{GHOST DR} - originally described in][]{ireland2018,hayes2022}, which was 
% modified by the \texttt{DRAGONS} team during the commissioning of GHOST.
%
%The data reduction was performed using a version of the {\texttt{DRAGONS}}\footnote{\href{https://github.com/GeminiDRSoftware/DRAGONS}{https://github.com/GeminiDRSoftware/DRAGONS}.} software package \citep{dragons,labrie2022} that includes the GHOST Data Reduction pipeline \citep[\texttt{GHOST DR};][]{ireland2018,hayes2022}. 
%
The reduction steps included bias/flat corrections, wavelength calibration, sky subtraction, barycentric correction, extraction of individual orders, and variance-weighted stitching of the spectral orders. The six individual exposures were combined using a simple mean without rejection. The signal-to-noise ratios per pixel achieved in selected regions of the spectrum are listed in Table~\ref{starinfo}. The colored panels on Figure~\ref{combo} show sections of the GHOST data (after normalization and radial velocity shift), highlighting absorption features of interest for the determination of stellar atmospheric parameters and chemical abundances, as described in Section~\ref{atmpars}.

\section{Atmospheric Parameters and Chemical Abundances}
\label{atmpars}

\subsection{Atmospheric Parameters}

The stellar atmospheric parameters (effective temperature -- \teff, surface gravity -- \logg, and metallicity -- \metal) for \rave\, were first calculated by \citet{Placco+2022} using the Gemini/GMOS data and the methods described therein. These parameters (\teff=4700~K, \logg=1.48, \metal=$-3.82$) were used to select \rave\, as a potential candidate for high-resolution spectroscopic follow-up.

In this work, the \teff\, for \rave\ was calculated from the color-\teff-\metal\, relations derived by \citet{mucciarelli21}. We used the same procedure outlined in \citet{roederer2018}, drawing 10$^5$ samples for magnitudes, reddening, and metallicity. The $G$, $BP$, and $RP$ magnitudes were retrieved from the third data release of the Gaia mission \citep[DR3;][]{gaia23dr3} and the $K$ magnitude from 2MASS \citep{skrutskie2006}. The final \teff=$4762\pm36$~K is the weighted mean of the median temperatures for each input color ($BP-RP$, $BP-G$, $G-RP$, $BP-K$, $RP-K$, and $G-K$).
The \logg\, was calculated using Equation~1 in \citet{roederer2018}, drawing 10$^5$ samples from the input parameters listed in Table~\ref{starinfo}. The final \logg=$1.58\pm0.11$ is taken as the median of those calculations with the uncertainty given by their standard deviation.

The metallicity was determined spectroscopically from the equivalent widths (EWs) of 104 \ion{Fe}{1} lines in the GHOST spectrum by fixing the \teff\ and \logg\ determined above. Table~\ref{eqw} lists the lines employed in this analysis, their measured equivalent widths, and the derived chemical abundances. The EWs were obtained by fitting Gaussian profiles to the observed absorption features using standard \texttt{IRAF}\footnote{IRAF was distributed by the National Optical Astronomy Observatory, which was managed by the Association of Universities for Research in Astronomy (AURA) under a cooperative agreement with the National Science Foundation.} routines, then \metal\, was calculated using the latest version of the \texttt{MOOG}\footnote{\href{https://github.com/alexji/moog17scat}{https://github.com/alexji/moog17scat}} code \citep{sneden1973}, employing one-dimensional plane-parallel model atmospheres with no overshooting \citep{castelli2004}, assuming local thermodynamic equilibrium (LTE). The microturbulent velocity ($\xi$) was determined by minimizing the trend between \ion{Fe}{1} abundances and their reduced equivalent width ($\log(\rm{EW}/\lambda$)). The final atmospheric parameters for \rave\, are listed in Table~\ref{starinfo}.

\begin{deluxetable}{lr@{}r@{}r@{}r@{}r@{}c@{}r}[!ht]
\tabletypesize{\tiny}
\tabletypesize{\footnotesize}
\tablewidth{0pc}
\tablecaption{\label{eqw} Atomic Data and Derived Abundances}
\tablehead{
\colhead{Ion}&
\colhead{$\lambda$}&
\colhead{$\chi$} &
\colhead{$\log\,gf$}&
\colhead{$EW$}&
\colhead{$\log\epsilon$\,(X)}&
\colhead{Ref.}&
\colhead{$\Delta$}\\
\colhead{}&
\colhead{({\AA})}&
\colhead{(eV)} &
\colhead{}&
\colhead{(m{\AA})}&
\colhead{}&
\colhead{}&
\colhead{NLTE}}
\startdata
CH             & 4313.00 & \nodata & \nodata & syn     & 4.83    &  1 & \nodata \\
\ion{Na}{1}    & 5889.95 & 0.00    &    0.11 & 142.63  &    3.57 &  1 & $-$0.37 \\
\ion{Na}{1}    & 5895.92 & 0.00    & $-$0.19 & 118.77  &    3.43 &  1 & $-$0.27 \\
\ion{Mg}{1}    & 3829.35 & 2.71    & $-$0.23 & 141.24  &    4.80 &  1 & 0.08 \\
\ion{Mg}{1}    & 3832.30 & 2.71    &    0.25 & 177.79  &    4.72 &  1 & 0.06 \\
\ion{Mg}{1}    & 3986.75 & 4.35    & $-$1.06 &  15.44  &    4.87 &  1 & \nodata \\
\ion{Mg}{1}    & 4167.27 & 4.35    & $-$0.74 &  18.56  &    4.64 &  1 & 0.13 \\
\ion{Mg}{1}    & 4702.99 & 4.33    & $-$0.44 &  34.70  &    4.65 &  1 & 0.18 \\
\ion{Mg}{1}    & 5172.68 & 2.71    & $-$0.36 & 156.21  &    4.79 &  1 & 0.05 \\
\ion{Mg}{1}    & 5183.60 & 2.72    & $-$0.17 & 177.67  &    4.84 &  1 & 0.04 \\
\nodata        & \nodata & \nodata & \nodata & \nodata & \nodata &    & \nodata \\ 
\enddata
    % \tablerefs{
    %  1: \citep{placco2021,placco2021a}
    %  2: \citet{nist};
    %  3: \citet{biemont2011};
    %  4: \citet{ljung2006};
    %  5: \citet{nist}, using HFS/IS from \citet{mcwilliam1998} when available;
    %  6: \citet{lawler2001}, using HFS from \citet{ivans2006} when available;
    %  7: \citet{lawler2009};
    %  8: \citet{li2007}, using HFS from \citet{sneden2009};
    %  9: \citet{denhartog2003}, using HFS/IS from \citet{roederer2008} when available;
    % 10: \citet{lawler2006}, using HFS/IS from \citet{roederer2008} when available;
    % 11: \citet{lawler2001}, using HFS/IS from \citet{ivans2006};
    % 12: \citet{denhartog2006};
    % 13: \citet{lawler2001b}, using HFS from \citet{lawler2001b};
    % 14: \citet{wickliffe2000};
    % 15: \citet{lawler2004}, using HFS from \citet{sneden2009};
    % 16: \citet{lawler2008};
    % 17: \citet{wickliffe1997}, using HFS from \citet{sneden2009};
    % 18: \citet{sneden2009} for log(gf) value and HFS/IS;
    % 19: \citet{lawler2007};
    % 20: \citet{quinet2006};
    % 21: \citet{xu2007}, using HFS/IS from \citet{cowan2005};
    % 22: \citet{nilsson2002}.
    % }
    \tablecomments{The complete list of absorption features and literature references are given in Table~\ref{eqwl}.}
    %\tablenotetext{a}{Using non-LTE corrections of \citet{lind2011}.}
\end{deluxetable}

\begin{deluxetable}{lcrrrcr}[!ht]
\tabletypesize{\small}
\tabletypesize{\footnotesize}
\tablewidth{0pc}
\tablecaption{LTE Abundances for Individual Species \label{abund}}
\tablehead{
\colhead{Species}                     & 
\colhead{$\log\epsilon_{\odot}$\,(X)} & 
\colhead{$\log\epsilon$\,(X)}         & 
\colhead{$\mbox{[X/H]}$}              & 
\colhead{$\mbox{[X/Fe]}$}             & 
\colhead{$\sigma$}                    & 
\colhead{$N$}}
\startdata
 \ion{C}{0}     &  8.43 &    4.83 & $-$3.60 & $-$0.21 &    0.10 &   1 \\
 \ion{C}{0}\vv  &  8.43 &    5.10 & $-$3.33 &    0.06 &    0.10 &   1 \\
 \ion{Na}{1}    &  6.24 &    3.50 & $-$2.74 &    0.65 &    0.07 &   2 \\ 
 \ion{Mg}{1}    &  7.60 &    4.75 & $-$2.85 &    0.54 &    0.07 &   9 \\
 \ion{Al}{1}    &  6.45 &    2.80 & $-$3.65 & $-$0.26 &    0.15 &   1 \\
 \ion{Si}{1}    &  7.51 &    4.76 & $-$2.75 &    0.64 &    0.15 &   1 \\
 \ion{Ca}{1}    &  6.34 &    3.36 & $-$2.98 &    0.41 &    0.08 &  11 \\
 \ion{Sc}{2}    &  3.15 & $-$0.02 & $-$3.17 &    0.22 &    0.05 &   7 \\
 \ion{Ti}{1}    &  4.95 &    1.70 & $-$3.25 &    0.14 &    0.05 &   7 \\
 \ion{Ti}{2}    &  4.95 &    1.94 & $-$3.01 &    0.38 &    0.09 &  20 \\
  \ion{V}{2}    &  3.93 &    0.97 & $-$2.96 &    0.43 &    0.02 &   2 \\
 \ion{Cr}{1}    &  5.64 &    1.79 & $-$3.85 & $-$0.46 &    0.10 &   3 \\
 \ion{Mn}{1}    &  5.43 &    1.51 & $-$3.92 & $-$0.53 &    0.09 &   3 \\
 \ion{Fe}{1}    &  7.50 &    4.11 & $-$3.39 &    0.00 &    0.12 & 104 \\
 \ion{Fe}{2}    &  7.50 &    4.19 & $-$3.31 &    0.08 &    0.04 &  11 \\
 \ion{Co}{1}    &  4.99 &    1.68 & $-$3.31 &    0.08 &    0.06 &   3 \\
 \ion{Ni}{1}    &  6.22 &    2.74 & $-$3.48 & $-$0.09 &    0.15 &   1 \\
 \ion{Zn}{1}    &  4.56 &    1.36 & $-$3.20 &    0.19 &    0.15 &   1 \\
 \ion{Sr}{2}    &  2.87 &    0.37 & $-$2.50 &    0.89 &    0.10 &   2 \\
  \ion{Y}{2}    &  2.21 & $-$0.74 & $-$2.95 &    0.44 &    0.04 &   6 \\
 \ion{Zr}{2}    &  2.58 & $-$0.20 & $-$2.77 &    0.62 &    0.04 &   4 \\
 \ion{Ba}{2}    &  2.18 &    0.04 & $-$2.14 &    1.25 &    0.10 &   3 \\
 \ion{La}{2}    &  1.10 & $-$1.01 & $-$2.11 &    1.28 &    0.06 &   8 \\
 \ion{Ce}{2}    &  1.58 & $-$0.63 & $-$2.21 &    1.18 &    0.06 &  10 \\
 \ion{Pr}{2}    &  0.72 & $-$1.15 & $-$1.87 &    1.52 &    0.02 &   3 \\
 \ion{Nd}{2}    &  1.42 & $-$0.58 & $-$2.00 &    1.39 &    0.04 &  22 \\
 \ion{Sm}{2}    &  0.96 & $-$0.92 & $-$1.88 &    1.51 &    0.08 &  17 \\
 \ion{Eu}{2}    &  0.52 & $-$1.25 & $-$1.77 &    1.62 &    0.05 &   8 \\
 \ion{Gd}{2}    &  1.07 & $-$0.74 & $-$1.81 &    1.58 &    0.04 &   9 \\
 \ion{Tb}{2}    &  0.30 & $-$1.35 & $-$1.65 &    1.74 &    0.20 &   1 \\
 \ion{Dy}{2}    &  1.10 & $-$0.47 & $-$1.57 &    1.82 &    0.12 &   9 \\
 \ion{Ho}{2}    &  0.48 & $-$1.33 & $-$1.81 &    1.58 &    0.08 &   3 \\
 \ion{Er}{2}    &  0.92 & $-$0.80 & $-$1.72 &    1.67 &    0.06 &   4 \\
 \ion{Tm}{2}    &  0.10 & $-$1.65 & $-$1.75 &    1.65 &    0.08 &   4 \\
 \ion{Yb}{2}    &  0.84 & $-$1.06 & $-$1.90 &    1.49 &    0.20 &   1 \\
 \ion{Hf}{2}    &  0.85 & $-$1.15 & $-$2.00 &    1.39 &    0.20 &   1 \\
 \ion{Os}{1}    &  1.40 & $-$0.21 & $-$1.61 &    1.78 &    0.04 &   2 \\
 \ion{Ir}{1}    &  1.38 & $-$0.35 & $-$1.73 &    1.66 &    0.20 &   1 \\
 \ion{Th}{2}    &  0.02 & $-$1.21 & $-$1.23 &    2.16 &    0.06 &   3 \\
\enddata
\tablenotetext{a}{Using carbon evolutionary corrections of \citet{placco2014c}.}
\end{deluxetable}

\begin{deluxetable}{@{}lrrrrr@{}}
\tabletypesize{\small}
\tabletypesize{\footnotesize}
\tablewidth{0pc}
\tablecaption{Example Systematic Abundance Uncertainties for \protect\rave \label{sys}}
\tablehead{
\colhead{Elem}&
\colhead{$\Delta$\teff}&
\colhead{$\Delta$\logg}&
\colhead{$\Delta\xi$}&
%\colhead{$\sigma/\sqrt{n}$}&
\colhead{$\sigma$}&
\colhead{$\sigma_{\rm tot}$}\\
\colhead{}&
\colhead{$+$150\,K}&
\colhead{$+$0.3 dex}&
\colhead{$+$0.3 km/s}&
\colhead{}&
\colhead{}}
\startdata
\ion{Na}{1} &    0.18 & $-$0.06 & $-$0.13 &    0.07 &    0.24 \\
\ion{Mg}{1} &    0.13 & $-$0.06 & $-$0.05 &    0.07 &    0.17 \\
%\ion{Al}{1} &   0.18 & $-$0.06 & $-$0.14 &    0.15 &    0.28 \\
%\ion{Si}{1} &   0.19 & $-$0.04 & $-$0.05 &    0.15 &    0.25 \\
\ion{Ca}{1} &    0.10 & $-$0.02 & $-$0.02 &    0.08 &    0.13 \\
\ion{Sc}{2} &    0.09 &    0.07 & $-$0.03 &    0.05 &    0.13 \\
\ion{Ti}{1} &    0.18 & $-$0.02 & $-$0.02 &    0.05 &    0.19 \\
\ion{Ti}{2} &    0.08 &    0.08 & $-$0.05 &    0.09 &    0.15 \\
%\ion{V}{2}  &   0.07 &    0.08 & $-$0.02 &    0.02 &    0.11 \\
\ion{Cr}{1} &    0.19 & $-$0.03 & $-$0.08 &    0.10 &    0.23 \\
\ion{Mn}{1} &    0.22 & $-$0.03 & $-$0.13 &    0.09 &    0.27 \\
\ion{Fe}{1} &    0.16 & $-$0.02 & $-$0.04 &    0.12 &    0.20 \\
\ion{Fe}{2} &    0.02 &    0.08 & $-$0.01 &    0.04 &    0.09 \\
\ion{Co}{1} &    0.20 & $-$0.02 & $-$0.07 &    0.06 &    0.22 \\
\ion{Ni}{1} &    0.17 & $-$0.01 & $-$0.02 &    0.15 &    0.23 \\
%\ion{Zn}{1} &   0.06 &    0.04 & $-$0.01 &    0.15 &    0.17 \\
\enddata
\end{deluxetable}

\begin{deluxetable}{lcrrrcr}[!ht]
\tabletypesize{\small}
\tabletypesize{\footnotesize}
\tablewidth{0pc}
\tablecaption{NLTE Abundances for Individual Species \label{abundn}}
\tablehead{
\colhead{Species}                     & 
\colhead{$\log\epsilon_{\odot}$\,(X)} & 
\colhead{$\log\epsilon$\,(X)}         & 
\colhead{$\mbox{[X/H]}$}              & 
\colhead{$\mbox{[X/Fe]}$}             & 
\colhead{$\sigma$}                    & 
\colhead{$N$}}
\startdata
 \ion{Na}{1}    &  6.24 &    3.18 & $-$3.06 &    0.15 &    0.03 &   2 \\ 
 \ion{Mg}{1}    &  7.60 &    4.83 & $-$2.77 &    0.44 &    0.05 &   8 \\
 \ion{Al}{1}    &  6.45 &    3.80 & $-$2.65 &    0.51 &    0.15 &   1 \\
 \ion{Si}{1}    &  7.51 &    4.79 & $-$2.72 &    0.51 &    0.15 &   1 \\
 \ion{Ca}{1}    &  6.34 &    3.64 & $-$2.70 &    0.41 &    0.07 &   9 \\
 \ion{Ti}{1}    &  4.95 &    2.30 & $-$2.65 &    0.56 &    0.06 &   6 \\
 \ion{Ti}{2}    &  4.95 &    2.04 & $-$2.91 &    0.30 &    0.10 &  19 \\
 \ion{Cr}{1}    &  5.64 &    2.49 & $-$3.15 &    0.06 &    0.06 &   3 \\
 \ion{Mn}{1}    &  5.43 &    1.83 & $-$3.60 & $-$0.39 &    0.08 &   3 \\
 \ion{Fe}{1}    &  7.50 &    4.29 & $-$3.21 &    0.00 &    0.13 & 102 \\
 \ion{Co}{1}    &  4.99 &    2.47 & $-$2.52 &    0.69 &    0.02 &   3 \\
\enddata
\tablecomments{The complete list of literature references for the NLTE corrections is given in Table~\ref{eqwl}.}
\end{deluxetable}

\subsection{Chemical Abundances}

The GHOST spectrum allowed for the detection of 308 absorption features for 36 elements, spanning the wavelength range $3694\leq\lambda(\textrm{\AA})\leq8807$. Abundances were determined from equivalent-width analysis and spectral synthesis, both using \texttt{MOOG}. These features and their atomic data are listed in Table~\ref{eqw}. Linelists for each abundance determination through spectral synthesis were generated using the \texttt{linemake} code\footnote{\href{https://github.com/vmplacco/linemake}{https://github.com/vmplacco/linemake}} \citep{placco2021,placco2021a}. Logarithmic abundances by number ($\log\epsilon$(X)) and abundance ratios (\abund{X}{H} and \xfe{X}), were calculated adopting the solar photospheric abundances ($\log\epsilon_{\odot}$\,(X)) from \citet{asplund2009}. The average abundances and the number of lines measured ($N$) for each element are given in Table~\ref{abund}. The $\sigma$ values are the standard error of the mean. For elements with only one line measured, the uncertainty was estimated by minimizing the residuals between the GHOST data and a set of synthetic spectra through visual inspection.

\begin{figure*}
 \includegraphics[width=1.0\linewidth]{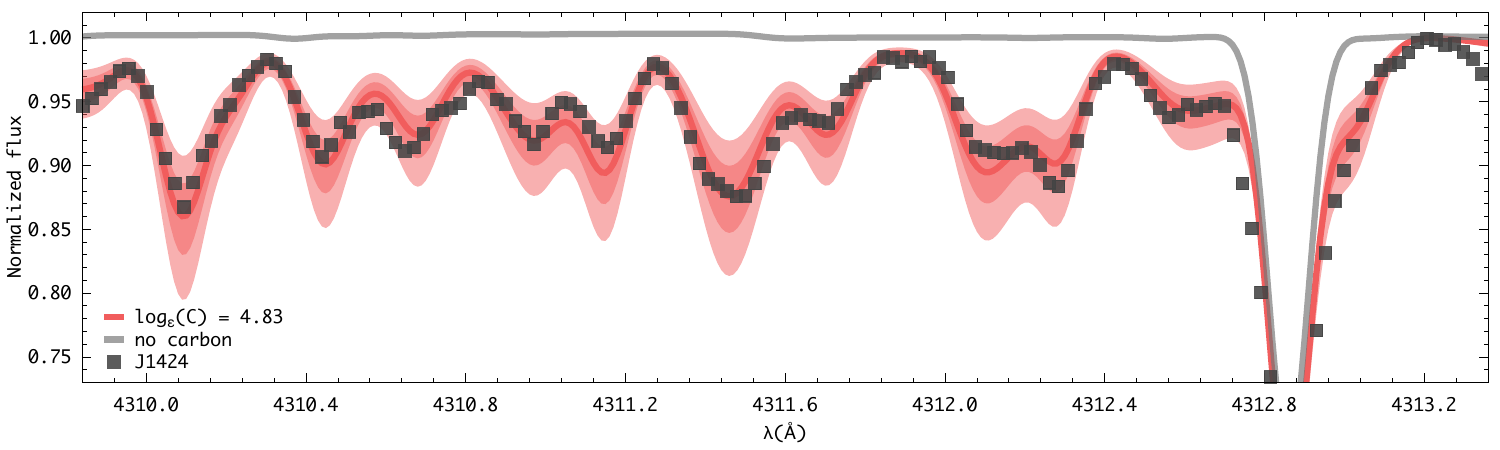}
     \caption{Spectral synthesis for the determination of the carbon abundance. The red solid line shows the best-fit synthesis and uncertainties ($\pm0.1$ and $\pm0.2$~dex - shaded regions) compared to the observed spectra (filled squares). Also shown is a synthetic spectrum after removing all contributions from carbon (gray line).}
 \label{synthesisc}
\end{figure*}

We have also quantified the systematic uncertainties due to changes in the atmospheric parameters for the elements with $6<Z\leq30$ with abundances determined by equivalent analysis only (see details below), following the prescription described in \citet{placco2013,placco2015b}.  Table~\ref{sys} shows the derived abundance variations when each atmospheric parameter is varied within the quoted uncertainties. Also listed is the total uncertainty for each element, calculated from the quadratic sum of the individual error estimates.  The adopted variations for the parameters are $+$150~K for \teff, $+$0.3~dex for \logg, and $+$0.3 km\,s$^{-1}$ for $\xi$. 

\subsubsection{From C to Zn}

Apart from C, Al, Si, V, and Zn, all the abundances for elements with $Z\leq30$ were measured from equivalent widths. The carbon abundance was determined from the CH G-band spectral synthesis, assuming $^{12}\rm{C}/^{13}\rm{C}=4$. Figure~\ref{synthesisc} shows the GHOST spectrum (filled squares) compared to the synthetic data. The red solid line shows the best-fit synthesis and the shaded regions at $\pm0.1$ and $\pm0.2$~dex are used to determine the uncertainty. Also shown is a synthetic spectrum after removing all contributions from carbon (gray line). The carbon depletion on the giant branch for \rave\, ($+0.27$~dex) was determined using the procedures described by \citet{placco2014c}\footnote{\href{https://vplacco.pythonanywhere.com/}{https://vplacco.pythonanywhere.com/}}.

\begin{figure*}
 \includegraphics[width=0.33\linewidth]{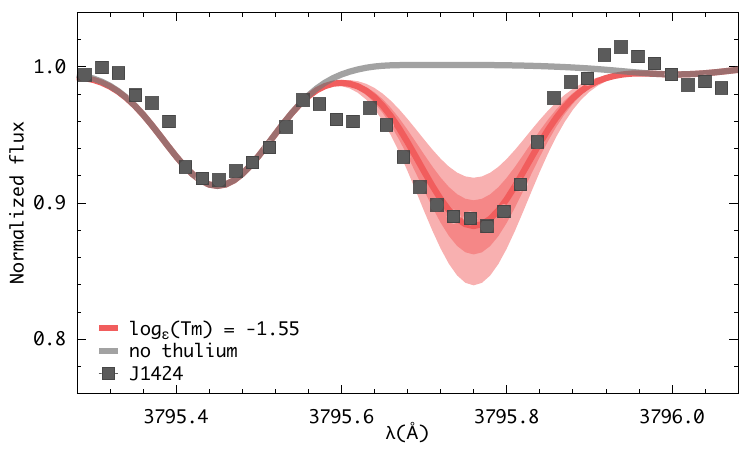}
 \includegraphics[width=0.33\linewidth]{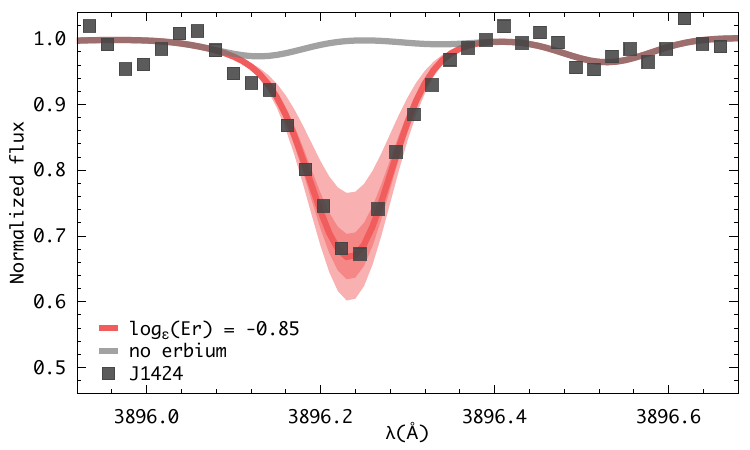}
 \includegraphics[width=0.33\linewidth]{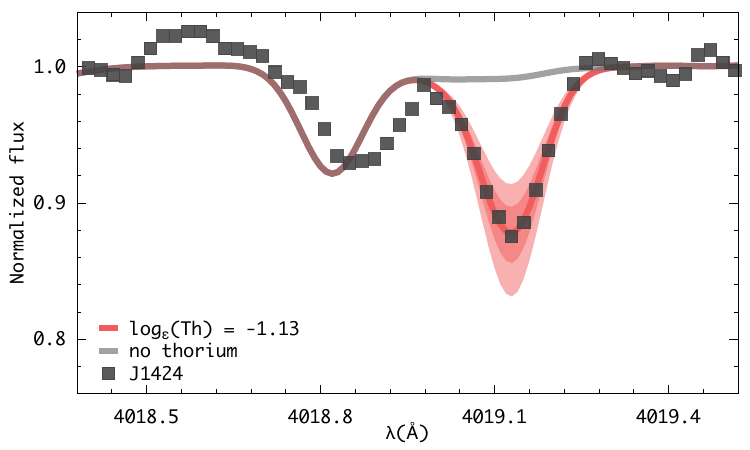}
 \includegraphics[width=0.33\linewidth]{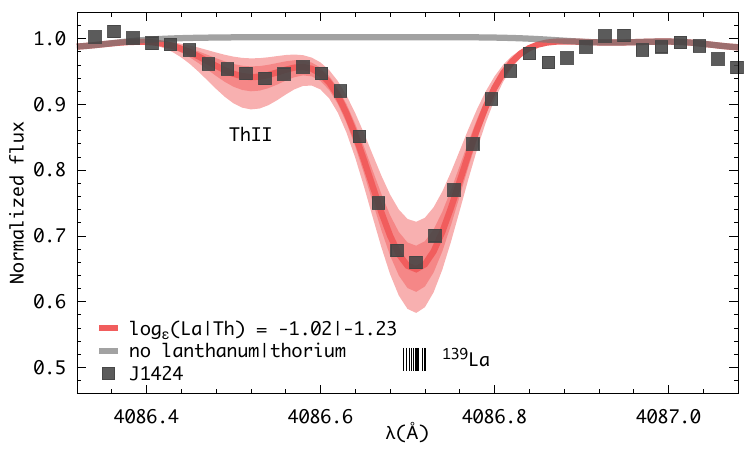}
 \includegraphics[width=0.33\linewidth]{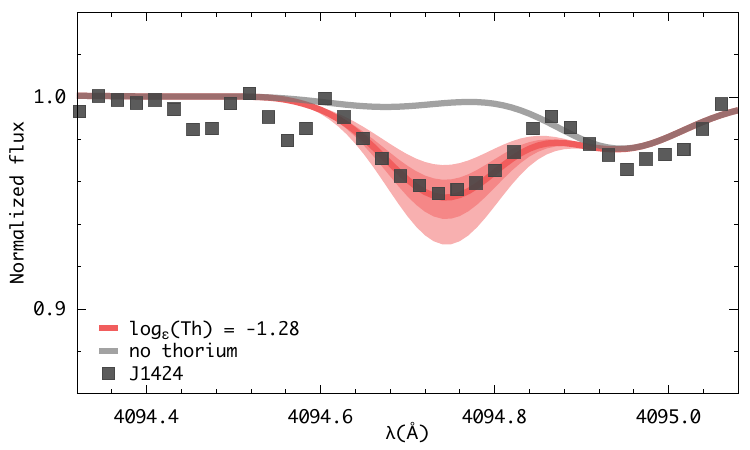}
 \includegraphics[width=0.33\linewidth]{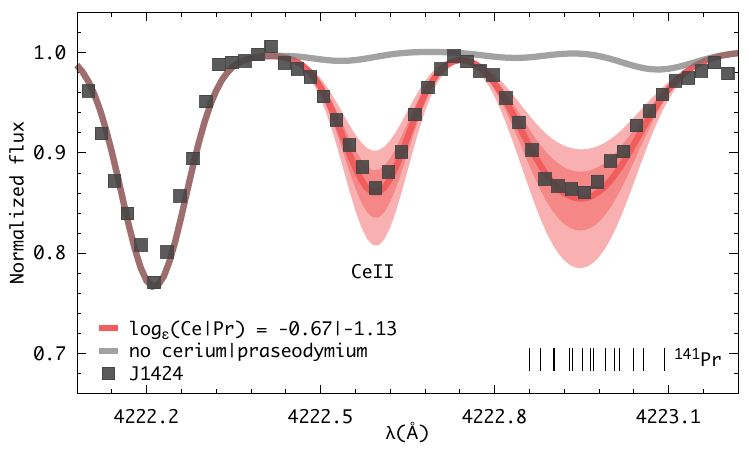}
 \includegraphics[width=0.33\linewidth]{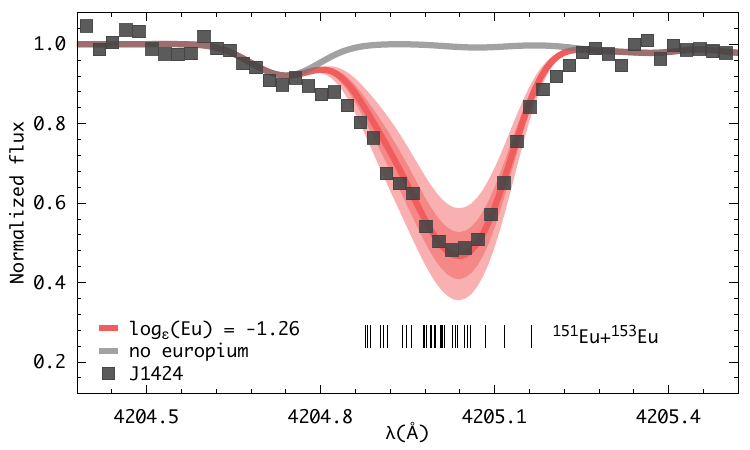}
 \includegraphics[width=0.33\linewidth]{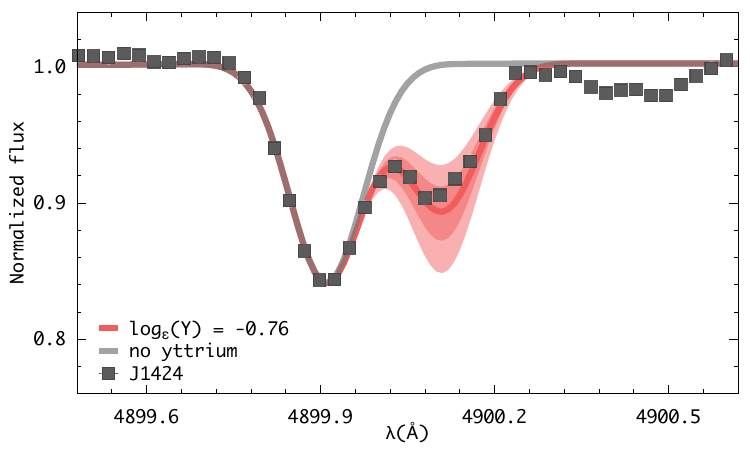}
 \includegraphics[width=0.33\linewidth]{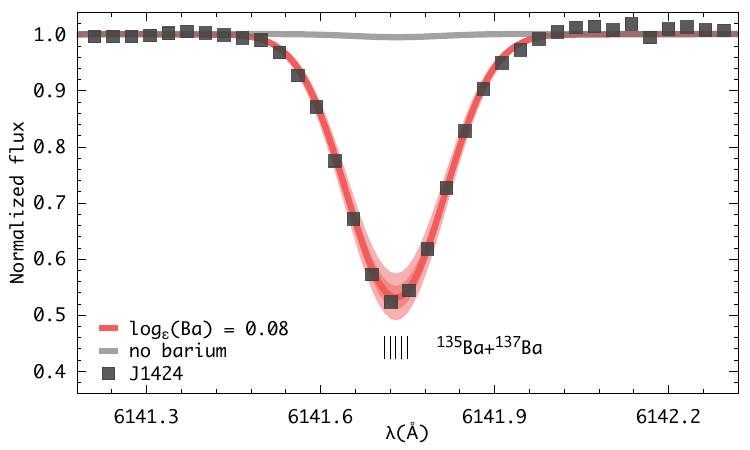}
     \caption{Same as Figure~\ref{synthesisc}, for the heavy elements highlighted in each panel.}
 \label{synthesisn}
\end{figure*}

For the remaining light elements, there is an overall good agreement among the abundances of individual lines for a given species, which can be seen from the small $\sigma$ values listed in Table~\ref{abund}. 
We have also obtained non-LTE (NLTE) corrections for 157 absorption features in the spectrum of \rave, using INSPECT\footnote{\href{http://www.inspect-stars.com/}{http://www.inspect-stars.com/}} (\ion{Na}{1}), \citet{nordlander2017b} (\ion{Al}{1}), and MPIA NLTE\footnote{\href{https://nlte.mpia.de/}{https://nlte.mpia.de/}} (\ion{Mg}{1}, \ion{Si}{1}, \ion{Ca}{1}, \ion{Ti}{1}, \ion{Ti}{2}, \ion{Cr}{1}, \ion{Mn}{1}, \ion{Fe}{1}, and \ion{Co}{1}). Literature references are given in Table~\ref{eqwl} along with the corrections for individual lines in the last column. Average NLTE abundances, abundance ratios, and $\sigma$ values are given in Table~\ref{abundn}. The average NLTE corrections range from $-0.32$ for \ion{Na}{1} to $+1.0$ for \ion{Al}{1}, with notably high corrections also for \ion{Cr}{1} and \ion{Co}{1} ($+0.79$ and $+0.70$, respectively).
%
%For the two \ion{Na}{1} D lines, we calculated the non-LTE corrections from \citet{lind2011}, using the INSPECT database, version 1.0\footnote{\href{http://www.inspect-stars.com/}{http://www.inspect-stars.com/}}. The discrepancy between the average abundances of \ion{Ti}{1} and \ion{Ti}{2} can also be attributed to non-LTE effects \citep{bergemann2011}. 
%
%
%
Due to the overall low metallicity (and low carbon abundance) of \rave, most lines have a well-defined continuum and are not blended with other species (see, for example, \ion{Mg}{1} and \ion{Ca}{2} in the lower panels of Figure~\ref{combo}).
Unless otherwise noted, we use the LTE abundances from Table~\ref{abund} for the remainder of this work.

\subsubsection{From Sr to Th}

\begin{figure*}
 \includegraphics[width=1.00\linewidth]{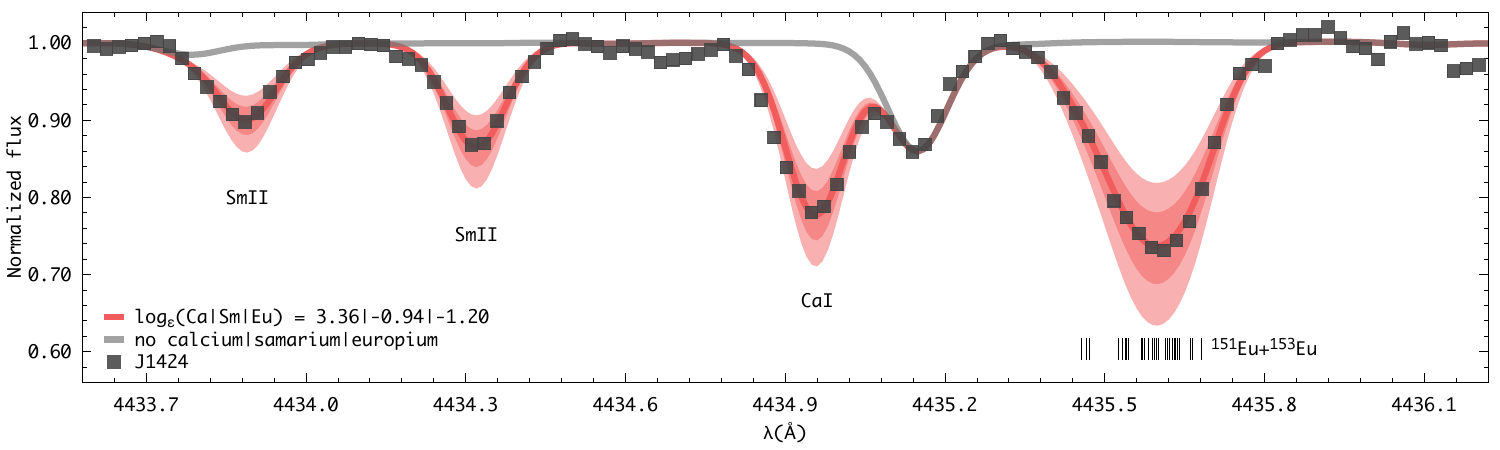}
     \caption{Same as Figure~\ref{synthesisc}, for samarium, calcium, and europium abundance determinations.}
 \label{synthesise}
\end{figure*}

The spectral synthesis of 121 absorption features was conducted for 21 chemical species with $Z\geq38$ and summarized in Table~\ref{abund}. Where appropriate, we accounted for line broadening by isotopic shifts and hyperfine splitting structure. For all syntheses, we fixed the abundances of carbon, iron, and the $^{12}$C/$^{13}$C ratio. We also used the $r$-process isotopic fractions from \citet{sneden2008} for specific elements, as described below. Figures~\ref{synthesisn} and \ref{synthesise} show the spectral synthesis for selected heavy elements. Symbols and lines have the same meaning as those shown in Figure~\ref{synthesisc}.

\paragraph{Strontium, yttrium, zirconium} For these first-peak elements, there is an excellent agreement between the abundances for individual lines. Both Sr lines ($\lambda$4077 and $\lambda$4215) were fit with the same abundance (\eps{Sr}=0.37) and the spread is small for the six Y lines (0.12~dex) and four Zr lines (0.10~dex). The synthesis for one of the Y lines is shown in Figure~\ref{synthesisn}. 

\paragraph{Barium, lanthanum} These second-peak elements have low $r$-process fractions \citep[Ba:15\%, La:25\% --][]{burris2000} in the Solar System.
For Ba, the strongest lines($\lambda$4554 and $\lambda$4934) appear saturated and were not considered in the analysis. The three Ba lines measured at redder wavelengths agree within 0.20~dex, with an average \eps{Ba}=$+0.04$. For La, the eight lines measured also agree within 0.20~dex, with an average of \eps{La}=$-1.01$. The syntheses for the Ba ($\lambda$6141) and La ($\lambda$4086 -- including hyperfine splitting) lines are shown in Figure~\ref{synthesisn}.

\paragraph{Cerium, praseodymium, neodymium, samarium} These elements have a large number of lines identified at wavelengths $\lambda\leq$4600\AA\, \citep[see][for a comprehensive list]{roederer2018}. In total, 52 lines were measured in the GHOST spectrum of \rave, with standard deviations $\sigma\leq0.08$. Figure~\ref{synthesise} shows the synthesis for two Sm lines and Figure~\ref{synthesisn} shows the synthesis for Ce and Pr (including hyperfine splitting).

\paragraph{Europium} This is one of the most widely used elements to indicate $r$-process nucleosynthesis and it is used to classify stars into various categories for heavy-element signatures \citep{frebel2018}. Eight lines were measured in the GHOST spectrum, ranging from $\lambda$3724 (\eps{Eu}=$-1.17$) to $\lambda$6645 (\eps{Eu}=$-1.23$). Two examples of Eu spectral synthesis are shown in Figure~\ref{synthesisn} ($\lambda$4205) and Figure~\ref{synthesise} ($\lambda$4435). In both cases, there is an overall good agreement between the observations (filled symbols) and the best synthetic fit (red lines). The final average is \eps{Eu}=$-1.25$ (\xfe{Eu}=$+1.62$).

\paragraph{Gadolinium, terbium, dysprosium, holmium, erbium, thulium, ytterbium, hafnium} These elements, with $64 \leq Z \leq 72$, are mostly formed by the $r$-process, according to the fractions in \citet{burris2000}. In total, 32 lines were measured within this group (most at $\lambda \leq$4000\AA, with only one feature for Tb ($\lambda$3874), Yb ($\lambda$3694), and Hf ($\lambda$4093). There were nine Dy lines measured with a somewhat high dispersion ($\sigma$=0.12~dex) and good agreement for Gd (9 lines -- $\sigma=0.04$), Ho (3 lines -- $\sigma=0.08$), Er (4 lines -- $\sigma=0.06$), and Tm (4 lines -- $\sigma=0.08$). The top panels of Figure~\ref{synthesisn} show the syntheses for Tm ($\lambda$3795) and Er ($\lambda3896$) lines.

\paragraph{Osmium, iridium} These third-peak elements are almost exclusively formed by the $r$-process in the Solar System \citep[Os:92\%, Ir:99\% --][]{burris2000} and also don't have many lines available for abundance determination in the spectral range of the GHOST data. The abundances for the two Os lines ($\lambda$4260 and $\lambda4420$) agree within 0.07~dex, with an average of \eps{Os}=$-0.21$. Only one Ir line was identified in \rave\, ($\lambda$3800), with an abundance of \eps{Ir}=$-0.35$.

\paragraph{Thorium} As a radioactive actinide with $Z=90$, Th is the second heaviest element with abundances measured in stellar spectra. For \rave, three lines were identified in the GHOST spectrum: $\lambda$4019 (\eps{Th}=$-1.13$), $\lambda$4086 (\eps{Th}=$-1.23$), and $\lambda$4094 (\eps{Th}=$-1.28$). Their spectral syntheses are shown in Figure~\ref{synthesisn}. For the $\lambda$4019 line, the abundances of C, Fe, Ni, Ce, and Nd were held constant using the average values in Table~\ref{abund}, and there appears to be a reduction artifact on the blue wing of the Th line. The La abundance was also held constant for the $\lambda$4086 synthesis. The GHOST spectrum was slightly smoothed (with a moving average of size 5 pixels) for the synthesis of the $\lambda$4094 line. The final average is \eps{Th}=$-1.21$.

\begin{figure*}
 \includegraphics[width=1\linewidth]{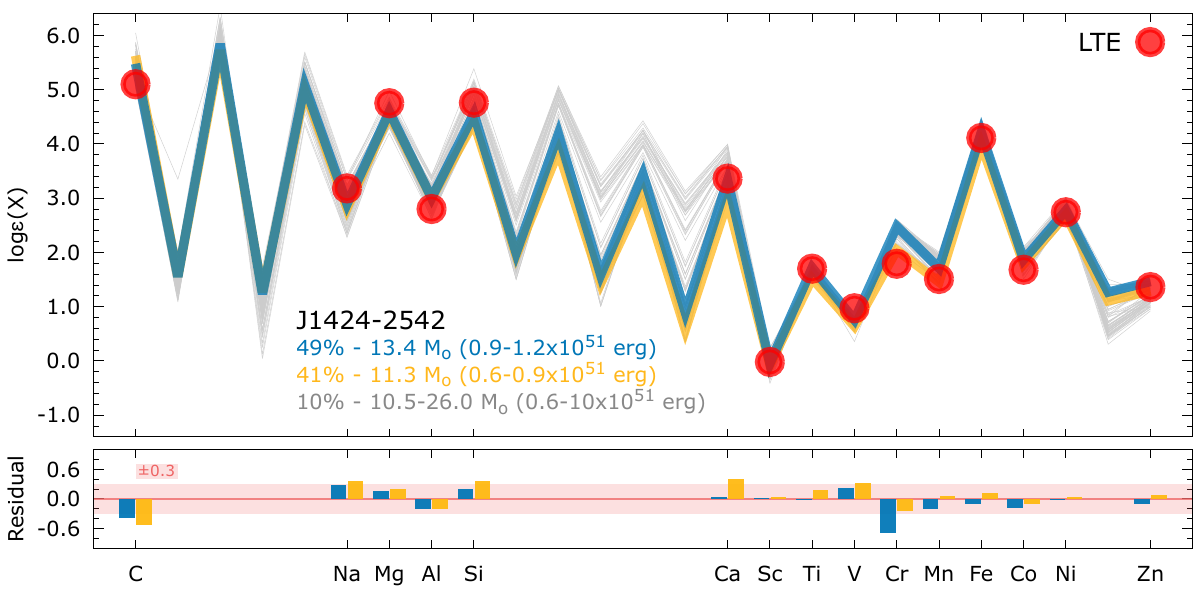}
     \caption{Upper panel: light-element chemical abundance pattern of \protect{\rave}, (filled circles, LTE) compared with yields from metal-free supernova models (solid lines). The labels show the progenitor mass and explosion energy of the models and their percentage occurrence among the 10,000 abundance pattern resamples of \protect{\rave}. Lower panel: residuals between observations and the two best-fit models. A $\pm$0.3~dex shaded area is shown for reference.}
 \label{starfit}
\end{figure*}

\section{The Chemo-dynamical Nature of \protect\rave}
\label{chemod}

In this section, we discuss the chemo-dynamical nature of \rave\, by comparing its chemical abundance pattern with Pop\,III supernova nucleosynthesis yields ($Z\leq30$), the $r$- and $s$- process solar fractions, and predictions from a simulation of neutron star mergers ($Z\geq38$). We also determine the mass, age, and orbit for \rave, in an attempt to constrain its formation history.

\subsection{The Light-element Abundance Pattern}
\label{light}

%cat mod/9486_abs.mod.res | awk '{print (($2-$3)^2)/sqrt($3^2)}' | datamash sum 1
%0.419944178
%cat mod/11293_abs.mod.res | awk '{print (($2-$3)^2)/sqrt($3^2)}' | datamash sum 1
%0.471574829
%cat mod/7703_abs.mod.res | awk '{print (($2-$3)^2)/sqrt($3^2)}' | datamash sum 1
%0.328403355
%cat mod/7969_abs.mod.res | awk '{print (($2-$3)^2)/sqrt($3^2)}' | datamash sum 1
%0.301740865

At \metal=$- 3.39$, \cfe=$+0.06$, and with enhancements in heavy elements, \rave\ most likely was formed from a gas cloud polluted by at least two progenitor populations. To corroborate that hypothesis, the \abund{Mg}{C} abundance ratio from \citet{hartwig2018} can be used as a diagnostic to distinguish between mono- and multi-enriched stars. For \rave, both the observed and natal values (\abund{Mg}{C}=$+0.75$ and $+0.48$, respectively) are consistent with the multi-enriched classification \citep[Figure 11 of][]{hartwig2018}.

Nonetheless, we can attempt to infer the main features of the progenitor population that enriched the gas cloud that formed \rave\, with elements from carbon to zinc. We modeled the light-element abundance signature of \rave\ by comparing it with the theoretical Pop\,III supernova nucleosynthesis yields\footnote{\href{http://starfit.org}{http://starfit.org}} from \citet{heger2010}.  These models predict the nucleosynthesis products of massive metal-free stars with pristine Big Bang nucleosynthesis initial composition, without mass loss and rotation throughout the evolution. The fallback models (\texttt{S4}) used in this work have masses from 10 to 100\,M$_\odot$ and explosion energies ranging from $0.3 \times 10^{51}$\,erg to $10\times 10^{51}$\,erg. The comparison between models and observations, as well as the $\chi^2$ matching algorithm, has already been applied to EMP stars in the literature \citep[][among others]{frebel2015b,roederer2016,placco2020} and provides important constraints on the progenitor population of second-generation stars.

\begin{figure*}
 \includegraphics[width=1\linewidth]{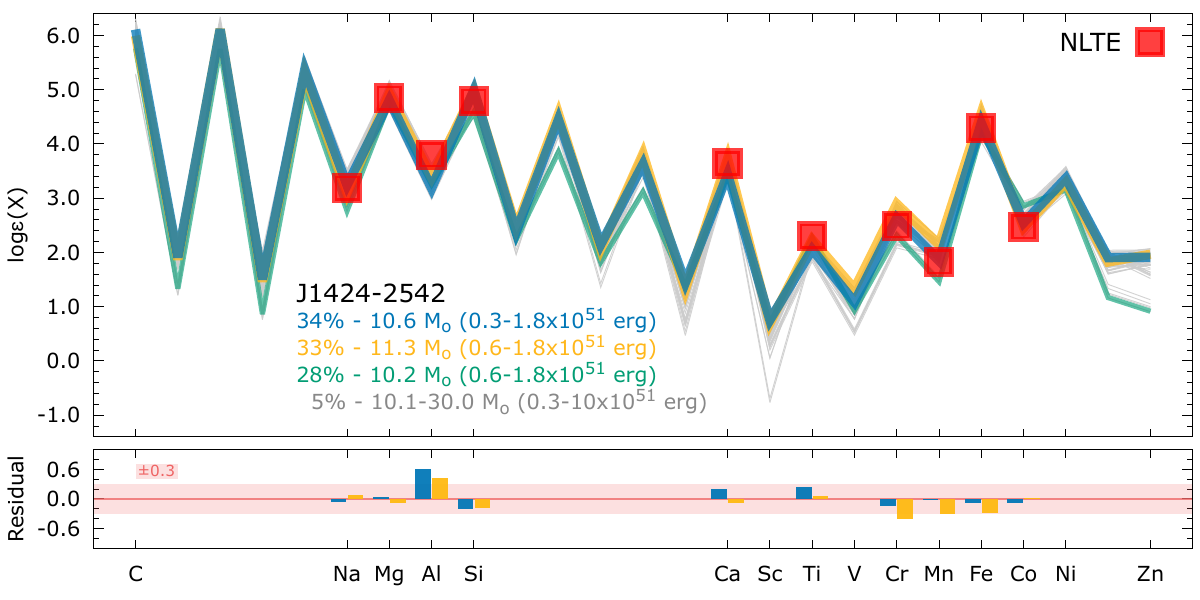}
     \caption{Same as Figure~\ref{starfit}, using the NLTE abundances listed in Table~\ref{abundn}.}
 \label{starfitn}
\end{figure*}

\begin{figure*}
 \includegraphics[width=1\linewidth]{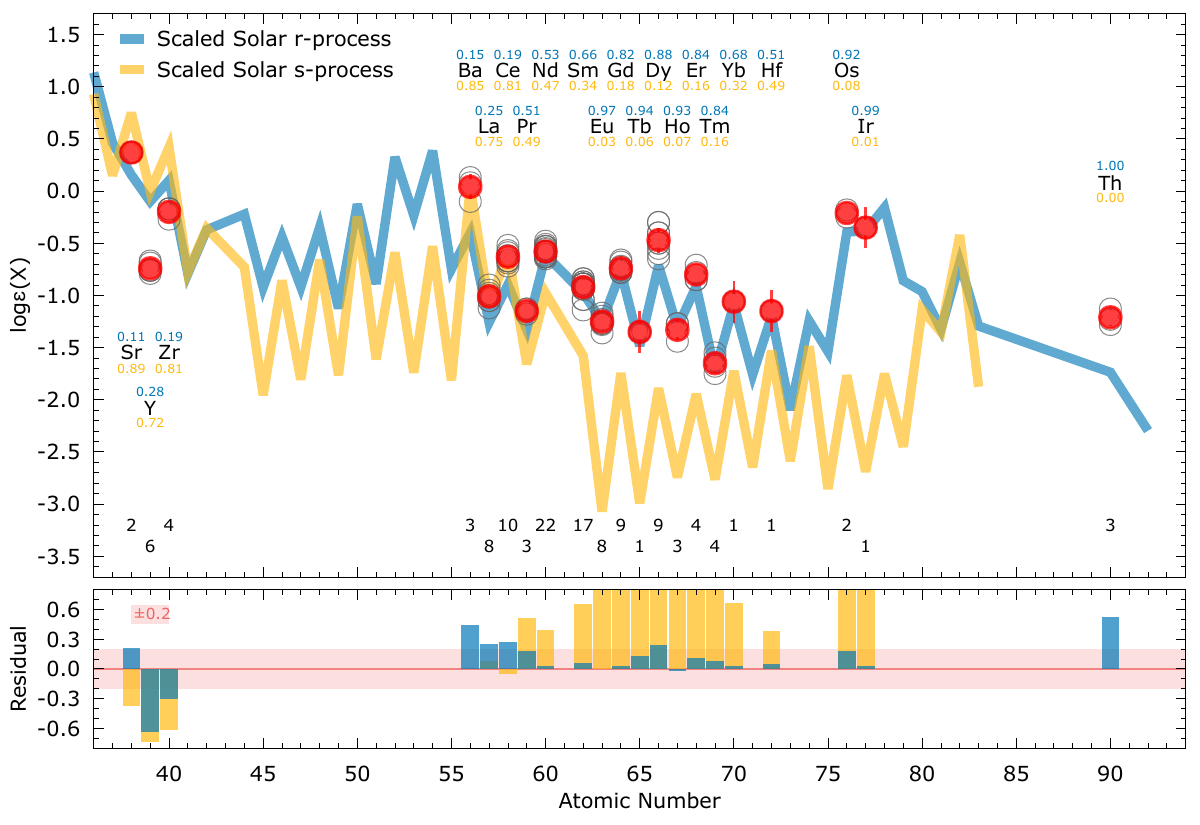}
     \caption{Upper panel: heavy-element chemical abundance pattern of \protect{\rave}, compared with the scaled solar system abundances. The r- and s-process contributions are calculated based on the fractions given by \citet{burris2000} and scaled to match the observed abundances of Eu and Ba, respectively. Also shown are the r- and s- process fractions for each element (top), as well as the number of lines identified for each element (bottom). Open circles show the individual abundances for all the lines measured in the GHOST spectrum. Lower panel: residuals between observations and the scaled solar system abundance patterns. A $\pm$0.2~dex shaded area is shown for reference.}
 \label{pattern}
\end{figure*}

Similar to \citet{placco2016b}, we created 10,000 abundance patterns for \rave, by re-sampling the $\log\epsilon (\mbox{X})$ and $\sigma$ values from Table~\ref{abund}. By determining the best-fit model for each re-sampled pattern using the LTE abundances, we found that 36 unique models provided an acceptable fit for at least 10 re-samples. The results of this exercise are shown in Figure~\ref{starfit}. In the upper panel, the filled circles show the chemical abundances for \rave\, and the lines represent the different models used for the fitting. The labels show the percentage occurrence for the most frequent models among the 10,000 runs. The bottom panel shows the residuals between observations and the three most frequent models.

The ``best-fit'' result found in 49\% of the re-samples is a model with 13.4\,M$_\odot$ [$0.9-1.2\times10^{51}$\,erg], followed by 11.3\,M$_\odot$ [$0.6-0.9\times10^{51}$\,erg] in 41\% of the re-samples. 
There is an overall good agreement between the two best-fit models and the observed abundances for \rave, with a somewhat large ($\geq+0.3$~dex) residual for carbon and chromium. It is interesting to note that, out of the 10,000 re-samples, about 90\% have their best-fit model for either $13.4\,M_\odot$ or $11.3\,M_\odot$ within a narrow range of explosion energies.

We repeated this exercise for the NLTE abundances in Table~\ref{abundn} and the results are shown in Figure~\ref{starfitn}. For the set of ten elements (as opposed to 15 in LTE), the most likely Pop. III characteristics are very similar to the LTE case, with a preference for lower masses and explosion energies. For 34\% of the re-samples, 10.6\,M$_\odot$ progenitors provide the best fit, followed by the 11.3\,M$_\odot$ (33\%) and 10.2\,M$_\odot$ (28\%) models, all with explosion energies within $0.3-1.8\times10^{51}$\,erg. Even though these results agree well with the LTE analysis, it is worth pointing out that carbon (and nitrogen) are key elements when comparing observations with the faint-SN models, as pointed out by \citet{placco2015}. Additional abundance determinations and NLTE corrections would help further constrain these models.

For both the LTE and NLTE abundance patterns, this exercise suggests that a progenitor star on the low-mass end of the SN grid with low explosion energy could be responsible for the light-element abundance pattern of \rave. This mass range and explosion energies are not consistent with the progenitor population suggested for stars with similar low carbon abundances: 30\,M$_\odot$ for \ump\, \citep{placco2021} and 20\,M$_\odot$ for AS0039 \citep{skuladottir2021}, both with explosion energy of $10\times10^{51}$\,erg. This may be a metallicity effect since these stars are in the \metal$<-4$ regime, so further exploration of the progenitor population of EMP stars would help better constrain their main characteristics.

\begin{figure*}[!ht]
 \includegraphics[width=1\linewidth]{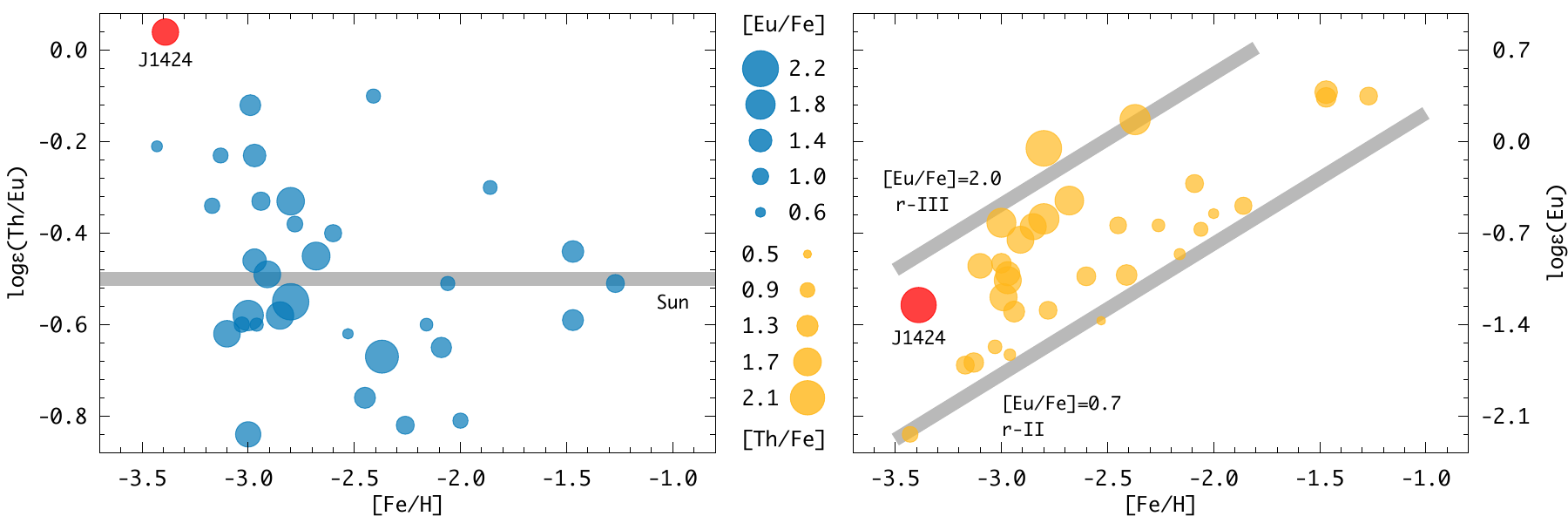}
     \caption{\eps{Th/Eu} (left panel) and \eps{Eu} (right panel) as a function of \metal\, for metal-poor stars in the literature with both Th and Eu measured (\metal$\leq-1.0$ and \xfe{Eu}$\geq+0.60$), compared to \rave. The point sizes are proportional to \xfe{Eu} (left) and \xfe{Th} (right), according to the labels between the panels. The horizontal solid gray line marks the solar \eps{Th/Eu} value and on the right panel, the limits for the $r$-II and $r$-III stars are shown. Individual references are given in Table~\ref{theutab}.}
 \label{theu}
\end{figure*}

\subsection{The Heavy-element Abundance Pattern}

With \xfe{Eu}=$+1.62$ and \abund{Ba}{Eu}=$-0.37$ abundance ratios, \rave\, is classified as an $r$-II metal-poor star \citep{frebel2018}, with a clear signature of the {\emph{main r-process}}. Its heavy-element abundance pattern, compared to the Solar System $s$-process (scaled to Ba) and $r$-process (scaled to Eu), is shown in the upper panel of Figure~\ref{pattern}. Filled circles are the average abundance for each element, while empty circles show the abundances for all the lines measured in the GHOST spectrum. Each label shows the element symbol and its $s$ and $r$ fractions, taken from \citet{burris2000}. Also shown are the number of lines used to calculate the average abundance for each element. The lower panel shows the residuals between observations and the scaled patterns. For reference, the red shaded area denotes the typical uncertainty ($\sim0.2$~dex) in the abundance measurements.

Sr, Y, and Zr agree with neither the scaled $s$ nor $r$ patterns for \rave. These elements are formed mainly by the $s$-process in the stars whose metals enriched the Sun. However, there are a number of possible formation channels for these light neutron-capture elements (dubbed as ``limited'' r-process), which could help explain their large variation, when compared with the normalized $r$-process patterns among low-metallicity stars \citep[see Table 2 and Figure 5 in][]{frebel2018}.
For Ba, La, and Ce, there is a clear over-production when compared to the scaled $r$-process pattern, which could suggest a contribution from the $s$-process to the observed abundance pattern of \rave. This contribution would be revealed by abundance ratios such as \abund{Ba}{Eu} and \abund{La}{Eu}, which are expected to be $\gtrsim 0.0$ if an $s$-process component is present \citep{roederer2010b,frebel2018}. For \rave, both ratios are consistent with the $r$-process expectation (and \abund{Ba}{Eu}=$-0.37$ and \abund{La}{Eu}=$-0.34$). 

In contrast, the abundances for elements from Pr to Ir well reproduce the normalized $r$-process pattern, mostly within 1-$\sigma$ (with the exception of Dy). Apart from those, thorium has a measured abundance that is over 0.5~dex higher than the normalized $r$-process pattern. This ``actinide boost'' phenomenon is shared by about a quarter of metal-poor stars with measurable Th (and U) and it could be evidence of either a contribution from a separate $r$-process event or small variations of neutron richness within the same type of $r$-process event that contributed to the abundance make up of \rave\, \citep{holmbeck2018,holmbeck2019}.

% \begin{figure}
%  \includegraphics[width=1\linewidth]{theu_new.pdf}
%      \caption{\eps{Th/Eu} as a function of \metal\, for metal-poor stars in the literature with both Th and Eu measured, compared to \rave. The point size is proportional to \xfe{Eu} and the solid gray line shows the solar value. Individual references are given in Table~\ref{theutab}.}
%  \label{theu}
% \end{figure}

Figure~\ref{theu} shows the heavy element abundance ratio \eps{Th/Eu} (left panel) and \eps{Eu} (right panel) as a function of \metal\, for stars in the literature\footnote{Taken from the JINAbase compilation \citep{jinabase}. Individual references are given in Table~\ref{theutab}.} with \metal$\leq-1.0$, \xfe{Eu}$\geq+0.6$, and both Th and Eu measured, compared to \rave. The point sizes are proportional to \xfe{Eu} (left) and \xfe{Th} (right). From the left panel, it is possible to see that \rave\, has the highest \eps{Th/Eu} within this group (well above the solar value - solid gray line) and the second lowest \metal, which corroborates with the hypothesis that it belongs to the ``actinide boost'' category and that its heavy elements have been produced by an $r$-process event without contributions from the $s$-process. The right panel also reveals that \rave\, has one of the highest \xfe{Th} ratios and the lowest metallicity among the $r$-II stars, and similar \xfe{Th} to the $r$-III star (\xfe{Eu}$\geq+2.0$) from \citet{cain2020}. In the following section, we present one possible scenario that can explain the heavy-element abundance pattern in \rave.

%awk '{print $13,$14}' residual.calc | awk '{print (($1-$2)^2)/2}' | datamash sum 1
%0.662297992
%awk '{print $6,$7}' residualNS.calc | awk '{print (($1-$2)^2)/2}' | datamash sum 1
%0.8237909033
%awk '$1>=56{print $13,$14}' residual.calc | awk '{print (($1-$2)^2)/2}' | datamash sum 1
%0.388460292
%awk '$1>=56{print $6,$7}' residualNS.calc | awk '{print (($1-$2)^2)/2}' | datamash sum 1
%0.4468763413

\vspace{1cm}

\subsection{Comparison with Yields from Neutron-Star Neutron-Star Merger Event}

Similarly to the exercise in Section~\ref{light} for the light elements, we explore the origin of the heavy elements in \rave\, made by the $r$-process.
% I'm struggling with this next sentence. Please help me reword!
Specifically, we use the analytic model of \citet{Holmbeck2021} to find which neutron star mergers can reproduce the observed abundance pattern of \rave.
This model predicts the total $r$-process yield for a neutron star merger using the neutron star masses and a nuclear equation of state (which determines their stellar radii) as input.
The total $r$-process yield is found by assuming a two-component ejecta scheme: a ``wind" and a ``dynamical" component.
The ejecta masses and compositions of the two components are calculated following the procedure and default model assumptions in \citet{Holmbeck2021}, namely that the ejecta masses of the wind and dynamical components follow the descriptions in \citet{Dietrich2020} and \citet{Krueger2020}, respectively.
We require the model output to match the relative light-to-heavy and actinide-to-heavy abundance features present in the abundance pattern of \rave, represented by the observational \eps{Zr/Dy} and \eps{Th/Dy} abundance ratios.
Using the nuclear equation of state proposed by \citet{Holmbeck2022}, we find that a 1.66--1.27 M$_\sun$ neutron star merger best reproduces these abundance ratios.

Including observational uncertainties, the neutron star masses can vary within $\pm$0.02 M$_\sun$ and still be able to match the elemental abundances in \rave. The model predicts median masses and lanthanide mass fractions of $m_{\rm disk} = 7.15_{-2.50}^{+2.96} \times 10^{-3}$~M$_\odot$ with $X_{\rm disk}^{\rm lan} = 0.050_{-0.017}^{+0.019}$ and $m_{\rm dyn} = 11.79_{-1.64}^{+1.40} \times 10^{-3}$~M$_\odot$ with $X_{\rm dyn}^{\rm lan} = 0.143_{-0.005}^{+0.005}$ for the disk and dynamical components, respectively.
% I say "high" too much here...
The model prefers a somewhat high total binary mass (2.93 M$_\odot$) and mass ratio ($M_1/M_2=1.31$) in order to minimize the light-to-heavy and maximize the actinide-to-heavy abundance ratios. The high total mass promotes a prompt collapse, maximizing the neutron-richness of the wind ejecta while also minimizing its total ejecta mass. This twofold effect serves to suppress the first $r$-process peak in favor of the heavy $r$-process elements: necessary in the present case of the relatively low first-peak abundances of \rave. At the same time, the high neutron star mass ratio promotes a high dynamical ejecta mass, which also serves to lower the light-to-heavy abundance ratio by diluting the wind ejecta with very neutron-rich dynamical ejecta that favors actinide production.

\begin{figure*}[!ht]
 \includegraphics[width=1\linewidth]{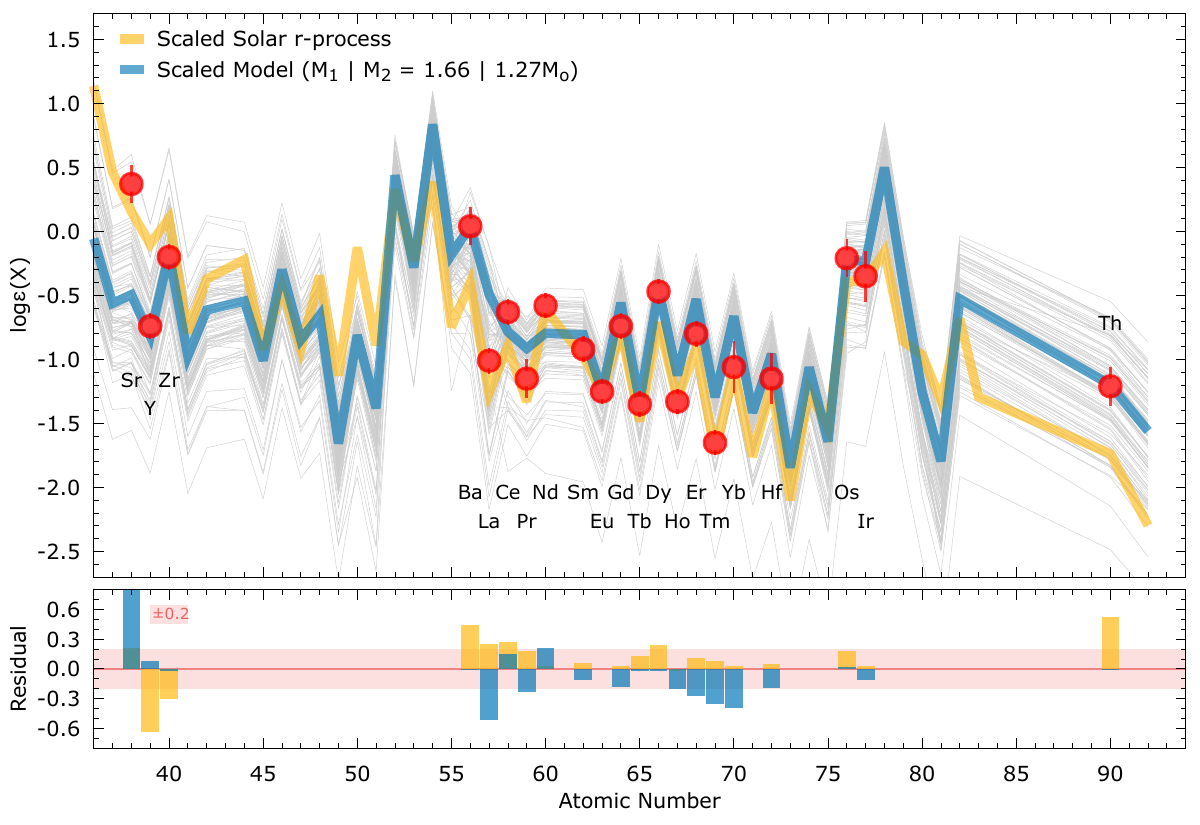}
 \caption{Upper panel: heavy-element chemical abundance pattern of \protect{\rave}, compared with the scaled solar system $r$-process (yellow) and the best-fit neutron star merger model (blue), normalized to match the observed europium abundance. Also shown (gray lines) are random realizations of the neutron star merger, see text for details. Lower panel: residuals between observations and scaled predictions. A $\pm$0.2~dex shaded area is shown for reference.}
 \label{nsmerger}
\end{figure*}

\begin{figure*}
 \includegraphics[width=1\linewidth]{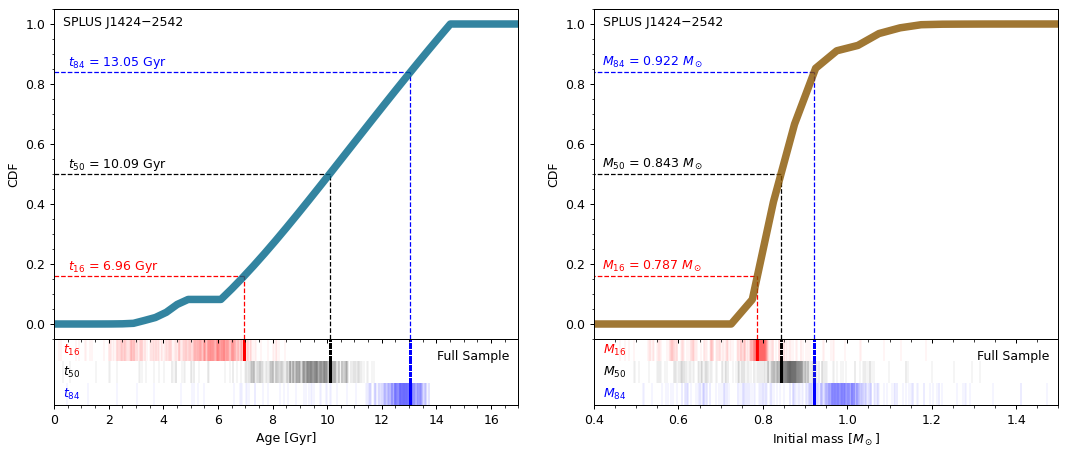}
     \caption{Age (left) and initial mass (right) cumulative distribution functions obtained for \rave\, using the Bayesian isochronal method described in \citet{AlmeidaFernandes+2023}. The dashed lines correspond to the 16th (red), 50th (black), and 84th (blue) percentiles used to characterize the parameters and their uncertainties. For comparison, the ticks in the bottom panels correspond to the 16th (red), 50th (black), and 84th (blue) percentiles for all 522 stars in the \citet{Placco+2022} sample.}
 \label{fig:age_mass_cdf}
\end{figure*}

Figure~\ref{nsmerger} shows the heavy-element abundance pattern of the best-fit neutron star merger model (blue) compared to \rave\, (red) and the scaled Solar $r$-process abundance pattern (yellow). The analytic model is not without its own uncertainties; also shown in Figure~\ref{nsmerger} are the chemical abundance patterns of 100 random realizations of a 1.66--1.27 M$_\sun$ neutron star merger (gray lines). These uncertainties reflect those of the analytic forms of the ejecta masses described in \citet{Dietrich2020} and \citet{Krueger2020} \citep[see][for details]{Holmbeck2022}.
% More comments about the model/figure?
%
Even though there are still some discrepancies between the theoretical predictions and observations (most notably for Sr, La, Tm, and Yb), this model can successfully reproduce the heavy element abundance pattern of \rave. Additional measurements from higher S/N spectra will help further constrain and refine the models.

\subsection{Age and Initial Mass}

In \citet{AlmeidaFernandes+2023}, the chemo-dynamical properties and ages of the 522 metal-poor candidates selected by \citet{Placco+2022}, which includes \rave, were analyzed. Below we discuss the parameters obtained for this particular star and the results are summarized in Table~\ref{starinfo}.

The age and initial mass of \rave\, were estimated through a Bayesian isochronal method using the MESA Isochrones \& Stellar Tracks \citep[MIST;][]{Dotter+2016}. Details of the process can be found in \citet{AlmeidaFernandes+2023}. In Figure \ref{fig:age_mass_cdf} we present the cumulative distribution function (CDF) for the age (left panel) and initial mass (right panel) for \rave. These parameters were estimated from the median of the distributions (black dashed lines), and the lower and upper limits as the 16th and 84th percentiles (red and blue dashed lines, respectively). For comparison, we also show the distribution of median ages and initial masses for all 522 stars in the \citet{Placco+2022} sample as black ticks in the bottom panels, as well as the distributions of 16th and 84th percentiles as red and blue ticks, respectively.

The CDF in the left panel of Figure \ref{fig:age_mass_cdf} shows that the estimated age for \rave\, is poorly constrained beyond 6 Gyr, i.e. the linear CDF corresponds to a very flat probability distribution at these ages. This CDF results in a very high age uncertainty, where the lower and upper limits differ from the median by about 3 Gyr. Nevertheless, the characterized median age of $10.09$ Gyr places \rave\, among the top 18\% oldest stars in the \cite{Placco+2022} sample. The CDF in the right panel shows that the initial mass of \rave\, can be much better constrained. The observed sub-solar mass of $0.843_{-0.056}^{+0.079}$ is consistent with the expectation for such an old and metal-poor star.

\subsection{Kinematical Parameters}

We used the photo-geometric distances provided by \citet{Bailer-Jones+2021}, and the proper motions and line-of-sight velocities of Gaia DR3 \citep{GaiaCollaboration+2022} to calculate the kinematical parameters of \rave. Its Heliocentric Galactic rectangular velocity vector corresponds to $(U, V, W) = (-93, -29.4, +46.4)$ km s$^{-1}$, resulting in a total velocity of $V_\mathrm{Tot} = 108.0$ km s$^{-1}$.  In cartesian galactocentric coordinates, its current position corresponds to $(X, Y, Z) = (2.61, -3.50, 4.22)$ kpc. Given its current position and total velocity, one can infer that \rave\, belongs to the Galactic halo.

\subsection{Galactic Orbit and Halo Substructure Membership}

\begin{figure}
 \includegraphics[width=1\linewidth]{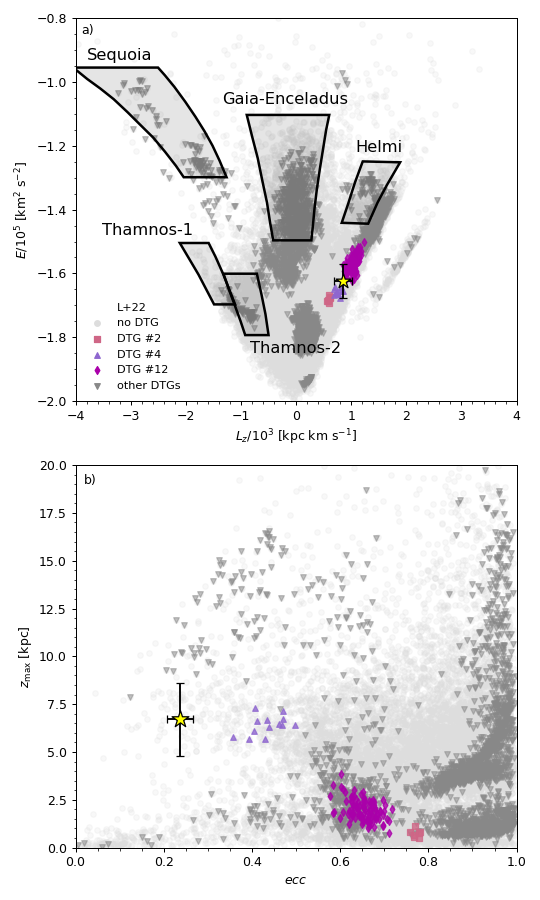}
 \caption{Top panel: comparison between the vertical component of the angular momentum and orbital energy of \rave\, (yellow star-shaped symbol) to those of known halo substructures (as defined by \citealp{Koppelman+2019}) and dynamically tagged groups (identified by \citealp{Lovdal+2022}). Bottom panel: eccentricity and maximum distance from the galactic plane for \rave\, and the stars in the \citet{Lovdal+2022} sample.}
 \label{fig:substructures_dtgs}
\end{figure}

The orbit of \rave\, was integrated in a \citet{McMillan2017} Galactic potential using the \texttt{galpy} package \citep{Bovy2015}. We adopted a galactocentric distance of $R_\odot = 8.21$ kpc, with corresponding rotation velocity of $V_\odot = 233.1$ km s$^{-1}$ \citep{McMillan2017}, and a Solar velocity of $(U_\odot, V_\odot, W_\odot) = (11.1, 12.24, 7.25)$ km s$^{-1}$ \citep{Schonrich+2010}. We estimated orbital parameters, such as apogalactic and perigalactic radius ($R_\mathrm{apo}$, $R_\mathrm{peri}$), maximum distance from the Galactic plane ($z_\mathrm{max}$), and orbital eccentricity ($ecc$). We also calculate the angular momentum, energy, and action angles, which allow us to investigate if \rave\, belongs to any known Halo substructure.

In Figure \ref{fig:substructures_dtgs} we compare the dynamical properties (top: $L_z$ vs. $E$; bottom: $ecc$ vs. $z_\mathrm{max}$) of \rave\, (yellow star-shaped symbol) with the parameters expected for different galactic substructures, as well as 67 dynamically tagged groups (DTGs). The uncertainties for \rave\, were computed from the standard deviation of the results from 5,000 orbital integrations produced using Monte Carlo re-sampling of the astrometry, distances, and radial velocities, taking into account the errors in each parameter. The shaded regions shown in the top panel correspond to the substructures of Sequoia, Thamnos-1 and Thamnos-2, Gaia-Sausage-Enceladus, and Helmi Stream, as defined by \citet{Koppelman+2019}. Given the observed differences in the vertical component of the angular momentum and in the energy, we can conclude that \rave\, does not share the same dynamical properties as any of the known halo major substructures.

We also compare the dynamical properties of \rave\, to those of 67 DTGs identified by \citet{Lovdal+2022} using data from Gaia EDR3 \citep{GaiaCollaboration+2021}. In Figure \ref{fig:substructures_dtgs} we include the sample of \citet{Lovdal+2022} (light-grey circles), and identify the stars that were assigned to any of the DTGs (grey inverted triangles). In the top panel, we highlight three DTGs that share similar $L_z$ and $E$ as \rave, labeled by \citet{Lovdal+2022} as DTGs 2 (pink squares), 4 (violet triangles), and 12 (magenta diamonds). However, as seen in the bottom panel, \rave\, does not share the same values of $ecc$ and $z_\mathrm{max}$ as DTGs 2 and 12. Stars in the DTG 4 have the same $z_\mathrm{max}$ as \rave, but the eccentricity is higher by about $0.2$. The differences between the dynamical properties of \rave\, and those of known halo substructures could be indicative that this star belongs to the in-situ halo population.

\vspace{0.5cm}

\section{Conclusions and Future Work}
\label{conclusion}

In this work, we presented the chemo-dynamical analysis of \rave, an $r$-process enhanced, actinide-boost star observed with the newly commissioned GHOST spectrograph at the Gemini South Telescope. By comparing the light- and heavy-element abundance patterns with yields from theoretical models, we speculate that the gas cloud from which \rave\, was formed must have been enriched by at least two progenitor populations, the supernova explosion from a metal-free 11.3--13.4\,M$_\odot$ star and the aftermath of a binary neutron star merger with masses 1.66\,M$_\odot$ and 1.27\,M$_\odot$. The mass ($0.843_{-0.056}^{+0.079}$\,M$_\odot$) and age ($10.09_{-3.12}^{+2.96}$ Gyr) for \rave\, are consistent with the proposed formation scenario and its kinematics do not connect it with any known structures in the Milky Way halo.
Further identification and spectroscopic follow-up of similar objects will help increase our understanding of the formation and chemical evolution of our Galaxy. In this context, GHOST will be a valuable resource for the astronomical community.

\vspace{2.0cm}

%The authors would like to thank the anonymous referee and members of the S-PLUS
%community (Guilherme Limberg, Leandro Beraldo e Silva, and Simone Daflon) who
%provided insightful comments on the manuscript.
%
The authors would like to thank Janice Lee, Bryan Miller, and the Gemini Observatory staff members for their contributions to the GHOST System Verification.
The work of V.M.P., K.C., E.D., J.H., V.K., S.X., R.D., M.G., D.H., P.P., C.Q., R.R., C.S., C.U., Z.H., D.J., K.L., B.M., G.P., S.R., and J.T. is supported by NOIRLab, which is managed by the Association of Universities for Research in Astronomy (AURA) under a cooperative agreement with the National Science Foundation.
F.A.-F. acknowledges funding for this work from FAPESP grants 2018/20977-2 and
2021/09468-1.
I.U.R.\ acknowledges support
from U.S.\ National Science Foundation (NSF) grants  
PHY~14-30152 (Physics Frontier Center/JINA-CEE),
AST~1815403/1815767, AST~2205847, and
the NASA Astrophysics Data Analysis Program,
grant 80NSSC21K0627.
K.A.V.\ thanks the National Sciences and Engineering Research Council of Canada for 
funding through the Discovery Grants and CREATE programs.
M.J. acknowledges partial support from the National Research Foundation (NRF) of Korea grant funded by the Ministry of Science and ICT (NRF-2021R1A2C1008679).
E.M. acknowledges funding from FAPEMIG under project number APQ-02493-22 and research productivity grant number 309829/2022-4 awarded by the CNPq, Brazil.
Based on observations obtained at the International Gemini Observatory (Program
IDs: GS-2021A-Q-419, GS-2023A-SV-101), a program of NSF’s
NOIRLab, which is managed by the Association of Universities for Research in
Astronomy (AURA) under a cooperative agreement with the National Science
Foundation. on behalf of the Gemini Observatory partnership: the National
Science Foundation (United States), National Research Council (Canada), Agencia
Nacional de Investigaci\'{o}n y Desarrollo (Chile), Ministerio de Ciencia,
Tecnolog\'{i}a e Innovaci\'{o}n (Argentina), Minist\'{e}rio da Ci\^{e}ncia,
Tecnologia, Inova\c{c}\~{o}es e Comunica\c{c}\~{o}es (Brazil), and Korea
Astronomy and Space Science Institute (Republic of Korea).
Data processed using DRAGONS (Data Reduction for Astronomy from Gemini Observatory North and South).
GHOST was built by a collaboration between Australian Astronomical Optics at Macquarie University, National Research Council Herzberg of Canada, and the Australian National University, and funded by the International Gemini partnership. The instrument scientist is Dr. Alan McConnachie at NRC, and the instrument team is also led by Dr. Gordon Robertson (at AAO), and Dr. Michael Ireland (at ANU). 
The authors would like to acknowledge the contributions of the GHOST instrument build team, the Gemini GHOST instrument team, the full SV team, and the rest of the Gemini operations team that were involved in making the SV observations a success.
The S-PLUS project, including the T80-South robotic telescope and the S-PLUS
scientific survey, was founded as a partnership between the Funda\c{c}\~{a}o de
Amparo \`{a} Pesquisa do Estado de S\~{a}o Paulo (FAPESP), the Observat\'{o}rio
Nacional (ON), the Federal University of Sergipe (UFS), and the Federal
University of Santa Catarina (UFSC), with important financial and practical
contributions from other collaborating institutes in Brazil, Chile (Universidad
de La Serena), and Spain (Centro de Estudios de F\'{\i}sica del Cosmos de
Arag\'{o}n, CEFCA). We further acknowledge financial support from the São Paulo
Research Foundation (FAPESP), the Brazilian National Research Council (CNPq),
the Coordination for the Improvement of Higher Education Personnel (CAPES), the
Carlos Chagas Filho Rio de Janeiro State Research Foundation (FAPERJ), and the
Brazilian Innovation Agency (FINEP).
The members of the S-PLUS collaboration are grateful for the contributions from
CTIO staff in helping in the construction, commissioning and maintenance of the
T80-South telescope and camera. We are also indebted to Rene Laporte, INPE, and
Keith Taylor for their important contributions to the project. From CEFCA, we
thank Antonio Mar\'{i}n-Franch for his invaluable contributions in the early
phases of the project, David Crist{\'o}bal-Hornillos and his team for their
help with the installation of the data reduction package \textsc{jype} version
0.9.9, C\'{e}sar \'{I}\~{n}iguez for providing 2D measurements of the filter
transmissions, and all other staff members for their support with various
aspects of the project.
IRAF was distributed by the National Optical Astronomy Observatory, which was 
managed by AURA under a cooperative agreement with the NSF.
This research has made use of NASA's Astrophysics Data System Bibliographic
Services; the arXiv pre-print server operated by Cornell University; the
{\texttt{SIMBAD}} database hosted by the Strasbourg Astronomical Data Center;
and the online Q\&A platform {\texttt{stackoverflow}}
(\href{http://stackoverflow.com/}{http://stackoverflow.com/}).

\software{
{\texttt{awk}}\,\citep{awk}, 
{\texttt{dustmaps}}\,\citep{green2018}, 
{\texttt{DRAGONS}}\,\citep{dragons,labrie2022}, 
{\texttt{gnuplot}}\,\citep{gnuplot}, 
{\texttt{IRAF}}\,\citep{tody1986,tody1993}, 
{\texttt{linemake}}\,\citep{placco2021,placco2021a},
%{\texttt{matplotlib}}\,\citep{matplotlib}, 
{\texttt{MOOG}}\,\citep{sneden1973},  
{\texttt{numpy}}\,\citep{numpy}, 
{\texttt{pandas}}\,\citep{pandas}, 
{\texttt{sed}}\,\citep{sed},
{\texttt{stilts}}\,\citep{stilts}.
}

\facilities{
Gemini:South (GMOS),
Gemini:South (GHOST)
}

\begin{deluxetable*}{lrrrr}
\tabletypesize{\small}
\tabletypesize{\footnotesize}
\tablewidth{0pc}
\tablecaption{Information for the literature comparison sample in Figure~\ref{theu}. \label{theutab}}
\tablehead{
\colhead{Star}&
\colhead{\metal}&
\colhead{\eps{Eu}}&
\colhead{\eps{Th}}&
\colhead{Reference}}
\startdata
LAMOST~J112456.61$+$453531.3 & $-$1.27 &   0.35 & $-$0.16 & \citet{XIN19} \\
HD~222925                    & $-$1.47 &   0.38 & $-$0.06 & \citet{roederer2008} \\
COS82                        & $-$1.47 &   0.34 & $-$0.25 & \citet{AOK07b} \\
RAVE~J093730.5$-$062655      & $-$1.86 & $-$0.49 & $-$0.79 & \citet{sakari2019} \\
2MJ1521$-$0607               & $-$2.00 & $-$0.55 & $-$1.36 & \citet{sakari2018} \\
BD$+$173248                  & $-$2.06 & $-$0.67 & $-$1.18 & \citet{COW02} \\
RAVE~J153830.9$-$180424      & $-$2.09 & $-$0.32 & $-$0.97 & \citet{sakari2018b} \\
%J1538$-$1804                & $-$2.09 & $-$0.30 & $-$0.95 & \citet{sakari2018} \\
HD~221170                    & $-$2.16 & $-$0.86 & $-$1.46 & \citet{IVA06} \\
2MJ2256$-$0719               & $-$2.26 & $-$0.64 & $-$1.46 & \citet{sakari2018} \\
2MASS~J21511791$-$1233417    & $-$2.37 &    0.17 & $-$0.50 & \citet{COH03} \\
2MASS~J15141890$+$0727028    & $-$2.41 & $-$1.02 & $-$1.12 & \citet{HON04} \\
J0246$-$1518                 & $-$2.45 & $-$0.64 & $-$1.40 & \citet{sakari2018} \\
HD~108317                    & $-$2.53 & $-$1.37 & $-$1.99 & \citet{ROE12b} \\
2MASS~J22310218$-$3238365    & $-$2.60 & $-$1.03 & $-$1.43 & \citet{HAY09} \\
2MASS~J00280692$-$2603042    & $-$2.68 & $-$0.45 & $-$0.90 & \citet{CHR04} \\
2MASS~J23303707$-$5626142    & $-$2.78 & $-$1.29 & $-$1.67 & \citet{MAS10} \\
J1521$-$3538                 & $-$2.80 & $-$0.05 & $-$0.60 & \citet{cain2018} \\
BD$-$16251                   & $-$2.80 & $-$0.59 & $-$0.92 & \citet{HON04} \\
CS~29497$-$004               & $-$2.85 & $-$0.65 & $-$1.23 & \citet{hill2017} \\
RAVE~J203843.2$-$002333      & $-$2.91 & $-$0.75 & $-$1.24 & \citet{placco2017} \\
2MASS~J22545856$-$4209193    & $-$2.94 & $-$1.30 & $-$1.63 & \citet{MAS14} \\
HD~115444                    & $-$2.96 & $-$1.63 & $-$2.23 & \citet{WES00} \\
J1432$-$4125                 & $-$2.97 & $-$1.01 & $-$1.47 & \citet{cain2018} \\
2MASS~J12213413$-$0328396    & $-$2.97 & $-$1.06 & $-$1.29 & \citet{HAY09} \\
2MASS~J09544277$+$5246414    & $-$2.99 & $-$1.19 & $-$1.31 & \citet{holmbeck2018} \\
TYC~5594$-$576$-$1           & $-$3.00 & $-$0.62 & $-$1.20 & \citet{FRE07b} \\
DES~J033523$-$540407         & $-$3.00 & $-$0.93 & $-$1.77 & \citet{ALE18} \\
J2005$-$3057                 & $-$3.03 & $-$1.57 & $-$2.17 & \citet{cain2018} \\
2MASS~J22170165$-$1639271    & $-$3.10 & $-$0.95 & $-$1.57 & \citet{SNE03} \\
2MASS~J01021585$-$6143458    & $-$3.13 & $-$1.69 & $-$1.92 & \citet{ROE14b} \\
LAMOST~J1109$+$0754          & $-$3.17 & $-$1.71 & $-$2.05 & \citet{MAR20} \\
\ravel                       & $-$3.39 & $-$1.25 & $-$1.21 & This work \\
2MASS~J23342669$-$2642140    & $-$3.43 & $-$2.24 & $-$2.45 & \citet{SIQ14} \\
\enddata
\end{deluxetable*}

\clearpage

\bibliographystyle{aasjournal}
%\bibliography{draft.bbl}

\twocolumngrid

\startlongtable

\begin{deluxetable}{lrrrrrcr}
\tabletypesize{\tiny}
\tablewidth{0pc}
\tablecaption{\label{eqwl} Atomic Data and Derived Abundances}
\tablehead{
\colhead{Ion}&
\colhead{$\lambda$}&
\colhead{$\chi$} &
\colhead{$\log\,gf$}&
\colhead{$EW$}&
\colhead{$\log\epsilon$\,(X)}&
\colhead{Ref.}&
\colhead{$\Delta$}\\
\colhead{}&
\colhead{({\AA})}&
\colhead{(eV)} &
\colhead{}&
\colhead{(m{\AA})}&
\colhead{}&
\colhead{}&
\colhead{NLTE}}
\startdata
CH & 4313.00 & \nodata & \nodata & syn & 4.83 & 1 & \nodata \\
\ion{Na}{1}  & 5889.95 & 0.00 &    0.11 & 142.63 &    3.57 &  1 & $-$0.37 \\
\ion{Na}{1}  & 5895.92 & 0.00 & $-$0.19 & 118.77 &    3.43 &  1 & $-$0.27 \\
\ion{Mg}{1}  & 3829.35 & 2.71 & $-$0.23 & 141.24 &    4.80 &  1 & 0.08 \\
\ion{Mg}{1}  & 3832.30 & 2.71 &    0.25 & 177.79 &    4.72 &  1 & 0.06 \\
\ion{Mg}{1}  & 3986.75 & 4.35 & $-$1.06 &  15.44 &    4.87 &  1 & \nodata \\
\ion{Mg}{1}  & 4167.27 & 4.35 & $-$0.74 &  18.56 &    4.64 &  1 & 0.13 \\
\ion{Mg}{1}  & 4702.99 & 4.33 & $-$0.44 &  34.70 &    4.65 &  1 & 0.18 \\
\ion{Mg}{1}  & 5172.68 & 2.71 & $-$0.36 & 156.21 &    4.79 &  1 & 0.05 \\
\ion{Mg}{1}  & 5183.60 & 2.72 & $-$0.17 & 177.67 &    4.84 &  1 & 0.04 \\
\ion{Mg}{1}  & 5528.40 & 4.35 & $-$0.55 &  33.23 &    4.73 &  1 & 0.16 \\
\ion{Mg}{1}  & 8806.76 & 4.35 & $-$0.14 &  69.25 &    4.75 &  1 & 0.01 \\
\ion{Al}{1}  & 3961.52 & 0.01 & $-$0.33 &    syn &    2.80 &  1 & 1.00 \\
\ion{Si}{1}  & 4102.94 & 1.91 & $-$3.34 &    syn &    4.76 &  1 & 0.03 \\
\ion{Ca}{1}  & 4283.01 & 1.89 & $-$0.20 &  25.88 &    3.28 &  1 & \nodata \\
\ion{Ca}{1}  & 4318.65 & 1.90 & $-$0.21 &  24.33 &    3.26 &  1 & \nodata \\
\ion{Ca}{1}  & 4425.44 & 1.88 & $-$0.41 &  21.26 &    3.35 &  1 & 0.28 \\
\ion{Ca}{1}  & 4454.78 & 1.90 &    0.26 &  49.42 &    3.30 &  1 & 0.29 \\
\ion{Ca}{1}  & 4455.89 & 1.90 & $-$0.55 &  14.52 &    3.31 &  1 & 0.24 \\
\ion{Ca}{1}  & 5588.76 & 2.52 &    0.30 &  19.62 &    3.28 &  1 & 0.39 \\
\ion{Ca}{1}  & 5857.45 & 2.93 &    0.17 &   9.36 &    3.50 &  1 & 0.21 \\
\ion{Ca}{1}  & 6102.72 & 1.88 & $-$0.81 &  16.11 &    3.52 &  1 & 0.28 \\
\ion{Ca}{1}  & 6122.22 & 1.89 & $-$0.33 &  28.82 &    3.38 &  1 & 0.27 \\
\ion{Ca}{1}  & 6162.17 & 1.90 & $-$0.11 &  39.86 &    3.39 &  1 & 0.26 \\
\ion{Ca}{1}  & 6439.07 & 2.52 &    0.33 &  24.00 &    3.34 &  1 & 0.24 \\
\ion{Sc}{2}  & 4320.73 & 0.60 & $-$0.25 &  54.81 & $-$0.10 &  1 & \nodata \\
\ion{Sc}{2}  & 4324.98 & 0.59 & $-$0.44 &  47.08 & $-$0.07 &  1 & \nodata \\
\ion{Sc}{2}  & 4400.39 & 0.60 & $-$0.54 &  43.58 & $-$0.03 &  1 & \nodata \\
\ion{Sc}{2}  & 4415.54 & 0.59 & $-$0.67 &  40.88 &    0.03 &  1 & \nodata \\
\ion{Sc}{2}  & 5031.01 & 1.36 & $-$0.40 &  12.94 & $-$0.05 &  1 & \nodata \\
\ion{Sc}{2}  & 5526.79 & 1.77 &    0.02 &  13.75 &    0.02 &  1 & \nodata \\
\ion{Sc}{2}  & 5657.91 & 1.51 & $-$0.60 &   8.03 &    0.06 &  1 & \nodata \\
\ion{Ti}{1}  & 3989.76 & 0.02 & $-$0.13 &  31.59 &    1.72 &  1 & 0.65 \\
\ion{Ti}{1}  & 3998.64 & 0.05 &    0.02 &  32.09 &    1.61 &  1 & 0.63 \\
\ion{Ti}{1}  & 4533.24 & 0.85 &    0.54 &  19.86 &    1.69 &  1 & 0.56 \\
\ion{Ti}{1}  & 4534.78 & 0.84 &    0.35 &  13.66 &    1.68 &  1 & 0.64 \\
\ion{Ti}{1}  & 4981.73 & 0.84 &    0.57 &  21.65 &    1.67 &  1 & 0.55 \\
\ion{Ti}{1}  & 4991.07 & 0.84 &    0.45 &  21.37 &    1.78 &  1 & 0.58 \\
\ion{Ti}{1}  & 5064.65 & 0.05 & $-$0.94 &   8.26 &    1.74 &  1 & \nodata \\
\ion{Ti}{2}  & 4337.91 & 1.08 & $-$0.96 &  75.85 &    2.07 &  1 & 0.10 \\
\ion{Ti}{2}  & 4394.06 & 1.22 & $-$1.77 &  28.54 &    2.02 &  1 & 0.08 \\
\ion{Ti}{2}  & 4395.03 & 1.08 & $-$0.54 &  93.18 &    2.12 &  1 & 0.09 \\
\ion{Ti}{2}  & 4395.84 & 1.24 & $-$1.93 &  17.54 &    1.93 &  1 & 0.08 \\
\ion{Ti}{2}  & 4399.77 & 1.24 & $-$1.20 &  46.68 &    1.84 &  1 & 0.15 \\
\ion{Ti}{2}  & 4417.71 & 1.17 & $-$1.19 &  59.84 &    2.01 &  1 & 0.13 \\
\ion{Ti}{2}  & 4418.33 & 1.24 & $-$1.99 &  12.44 &    1.81 &  1 & 0.13 \\
\ion{Ti}{2}  & 4443.80 & 1.08 & $-$0.71 &  82.36 &    1.97 &  1 & 0.01 \\
\ion{Ti}{2}  & 4450.48 & 1.08 & $-$1.52 &  46.20 &    1.96 &  1 & 0.01 \\
\ion{Ti}{2}  & 4464.45 & 1.16 & $-$1.81 &  24.12 &    1.89 &  1 & \nodata \\
\ion{Ti}{2}  & 4501.27 & 1.12 & $-$0.77 &  78.77 &    1.97 &  1 & 0.11 \\
\ion{Ti}{2}  & 4533.96 & 1.24 & $-$0.53 &  83.73 &    2.00 &  1 & 0.17 \\
\ion{Ti}{2}  & 4571.97 & 1.57 & $-$0.31 &  68.26 &    1.78 &  1 & 0.04 \\
\ion{Ti}{2}  & 4657.20 & 1.24 & $-$2.29 &   9.36 &    1.95 &  1 & 0.16 \\
\ion{Ti}{2}  & 5129.16 & 1.89 & $-$1.34 &  15.02 &    1.97 &  1 & 0.06 \\
\ion{Ti}{2}  & 5188.69 & 1.58 & $-$1.05 &  33.45 &    1.77 &  1 & 0.09 \\
\ion{Ti}{2}  & 5226.54 & 1.57 & $-$1.26 &  29.02 &    1.87 &  1 & 0.10 \\
\ion{Ti}{2}  & 5336.79 & 1.58 & $-$1.60 &  17.48 &    1.93 &  1 & 0.09 \\
\ion{Ti}{2}  & 5381.02 & 1.57 & $-$1.97 &  10.13 &    2.01 &  1 & 0.10 \\
\ion{Ti}{2}  & 5418.77 & 1.58 & $-$2.13 &   7.11 &    2.01 &  1 & 0.10 \\
\ion{V}{2}   & 3951.96 & 1.48 & $-$0.73 &    syn &    0.98 &  1 & \nodata \\
\ion{V}{2}   & 4005.70 & 1.82 & $-$0.45 &    syn &    0.96 &  1 & \nodata \\
\ion{Cr}{1}  & 4274.80 & 0.00 & $-$0.22 &  66.04 &    1.68 &  1 & 0.74 \\
\ion{Cr}{1}  & 4289.72 & 0.00 & $-$0.37 &  63.60 &    1.77 &  1 & 0.75 \\
\ion{Cr}{1}  & 5206.04 & 0.94 &    0.02 &  41.00 &    1.93 &  1 & 0.60 \\
\ion{Mn}{1}  & 4030.75 & 0.00 & $-$0.50 &  77.78 &    1.62 &  1 & 0.29 \\
\ion{Mn}{1}  & 4033.06 & 0.00 & $-$0.65 &  69.10 &    1.52 &  1 & 0.31 \\
\ion{Mn}{1}  & 4034.48 & 0.00 & $-$0.84 &  57.13 &    1.40 &  1 & 0.34 \\
\ion{Fe}{1}  & 3727.62 & 0.96 & $-$0.61 & 111.84 &    4.15 &  1 & 0.16 \\
\ion{Fe}{1}  & 3742.62 & 2.94 & $-$0.81 &  18.58 &    4.18 &  1 & 0.11 \\
\ion{Fe}{1}  & 3758.23 & 0.96 & $-$0.01 & 148.90 &    4.12 &  1 & 0.08 \\
\ion{Fe}{1}  & 3763.79 & 0.99 & $-$0.22 & 126.22 &    4.06 &  1 & 0.12 \\
\ion{Fe}{1}  & 3805.34 & 3.30 &    0.31 &  52.41 &    4.27 &  1 & 0.24 \\
\ion{Fe}{1}  & 3808.73 & 2.56 & $-$1.17 &  24.17 &    4.25 &  1 & 0.23 \\
\ion{Fe}{1}  & 3815.84 & 1.48 &    0.24 & 128.59 &    4.15 &  1 & 0.14 \\
\ion{Fe}{1}  & 3840.44 & 0.99 & $-$0.50 & 112.01 &    3.99 &  1 & 0.07 \\
\ion{Fe}{1}  & 3841.05 & 1.61 & $-$0.04 & 107.23 &    4.15 &  1 & 0.07 \\
\ion{Fe}{1}  & 3902.95 & 1.56 & $-$0.44 &  93.92 &    4.12 &  1 & 0.16 \\
\ion{Fe}{1}  & 3949.95 & 2.18 & $-$1.25 &  32.91 &    4.08 &  1 & 0.21 \\
\ion{Fe}{1}  & 3977.74 & 2.20 & $-$1.12 &  32.75 &    3.96 &  1 & 0.21 \\
\ion{Fe}{1}  & 4005.24 & 1.56 & $-$0.58 &  94.43 &    4.25 &  1 & 0.16 \\
\ion{Fe}{1}  & 4067.98 & 3.21 & $-$0.53 &  16.14 &    4.11 &  1 & 0.14 \\
\ion{Fe}{1}  & 4071.74 & 1.61 & $-$0.01 & 115.10 &    4.23 &  1 & 0.12 \\
\ion{Fe}{1}  & 4134.68 & 2.83 & $-$0.65 &  26.62 &    4.07 &  1 & 0.22 \\
\ion{Fe}{1}  & 4143.87 & 1.56 & $-$0.51 &  98.42 &    4.24 &  1 & 0.18 \\
\ion{Fe}{1}  & 4147.67 & 1.49 & $-$2.07 &  38.12 &    4.17 &  1 & 0.01 \\
\ion{Fe}{1}  & 4154.50 & 2.83 & $-$0.69 &  20.48 &    3.96 &  1 & 0.22 \\
\ion{Fe}{1}  & 4156.80 & 2.83 & $-$0.81 &  18.45 &    4.02 &  1 & 0.07 \\
\ion{Fe}{1}  & 4157.78 & 3.42 & $-$0.40 &  11.37 &    4.04 &  1 & 0.06 \\
\ion{Fe}{1}  & 4174.91 & 0.91 & $-$2.94 &  36.20 &    4.30 &  1 & 0.21 \\
\ion{Fe}{1}  & 4181.76 & 2.83 & $-$0.37 &  29.50 &    3.86 &  1 & 0.22 \\
\ion{Fe}{1}  & 4184.89 & 2.83 & $-$0.87 &  13.73 &    3.92 &  1 & 0.22 \\
\ion{Fe}{1}  & 4187.04 & 2.45 & $-$0.56 &  48.93 &    4.01 &  1 & 0.19 \\
\ion{Fe}{1}  & 4187.80 & 2.42 & $-$0.51 &  48.42 &    3.91 &  1 & 0.19 \\
\ion{Fe}{1}  & 4191.43 & 2.47 & $-$0.67 &  42.14 &    4.00 &  1 & 0.19 \\
\ion{Fe}{1}  & 4195.33 & 3.33 & $-$0.49 &  13.13 &    4.09 &  1 & 0.21 \\
\ion{Fe}{1}  & 4199.10 & 3.05 &    0.16 &  45.02 &    3.92 &  1 & 0.22 \\
\ion{Fe}{1}  & 4222.21 & 2.45 & $-$0.91 &  29.00 &    3.94 &  1 & 0.19 \\
\ion{Fe}{1}  & 4227.43 & 3.33 &    0.27 &  34.40 &    3.90 &  1 & 0.25 \\
\ion{Fe}{1}  & 4233.60 & 2.48 & $-$0.60 &  41.57 &    3.92 &  1 & 0.19 \\
\ion{Fe}{1}  & 4250.12 & 2.47 & $-$0.38 &  58.78 &    4.05 &  1 & 0.17 \\
\ion{Fe}{1}  & 4250.79 & 1.56 & $-$0.71 &  93.16 &    4.26 &  1 & 0.17 \\
\ion{Fe}{1}  & 4260.47 & 2.40 &    0.08 &  80.60 &    4.05 &  1 & 0.16 \\
\ion{Fe}{1}  & 4271.15 & 2.45 & $-$0.34 &  61.22 &    4.04 &  1 & 0.12 \\
\ion{Fe}{1}  & 4282.40 & 2.18 & $-$0.78 &  59.13 &    4.13 &  1 & 0.21 \\
\ion{Fe}{1}  & 4325.76 & 1.61 &    0.01 & 116.64 &    4.17 &  1 & 0.07 \\
\ion{Fe}{1}  & 4352.73 & 2.22 & $-$1.29 &  28.01 &    4.01 &  1 & 0.20 \\
\ion{Fe}{1}  & 4415.12 & 1.61 & $-$0.62 &  95.25 &    4.23 &  1 & 0.13 \\
\ion{Fe}{1}  & 4430.61 & 2.22 & $-$1.73 &  19.76 &    4.24 &  1 & 0.21 \\
\ion{Fe}{1}  & 4442.34 & 2.20 & $-$1.23 &  33.59 &    4.04 &  1 & 0.21 \\
\ion{Fe}{1}  & 4443.19 & 2.86 & $-$1.04 &  15.25 &    4.16 &  1 & 0.22 \\
\ion{Fe}{1}  & 4447.72 & 2.22 & $-$1.36 &  29.19 &    4.10 &  1 & 0.21 \\
\ion{Fe}{1}  & 4466.55 & 2.83 & $-$0.60 &  36.48 &    4.21 &  1 & 0.23 \\
\ion{Fe}{1}  & 4494.56 & 2.20 & $-$1.14 &  37.73 &    4.03 &  1 & 0.22 \\
\ion{Fe}{1}  & 4531.15 & 1.48 & $-$2.10 &  37.93 &    4.14 &  1 & 0.21 \\
\ion{Fe}{1}  & 4592.65 & 1.56 & $-$2.46 &  22.11 &    4.24 &  1 & 0.21 \\
\ion{Fe}{1}  & 4602.94 & 1.49 & $-$2.21 &  38.08 &    4.26 &  1 & 0.03 \\
\ion{Fe}{1}  & 4733.59 & 1.49 & $-$2.99 &  10.86 &    4.30 &  1 & 0.21 \\
\ion{Fe}{1}  & 4736.77 & 3.21 & $-$0.67 &  14.26 &    4.14 &  1 & 0.21 \\
\ion{Fe}{1}  & 4871.32 & 2.87 & $-$0.34 &  39.44 &    4.02 &  1 & 0.20 \\
\ion{Fe}{1}  & 4872.14 & 2.88 & $-$0.57 &  25.60 &    3.97 &  1 & 0.20 \\
\ion{Fe}{1}  & 4890.76 & 2.88 & $-$0.38 &  37.95 &    4.04 &  1 & 0.20 \\
\ion{Fe}{1}  & 4891.49 & 2.85 & $-$0.11 &  47.50 &    3.93 &  1 & 0.20 \\
\ion{Fe}{1}  & 4918.99 & 2.86 & $-$0.34 &  33.17 &    3.88 &  1 & 0.20 \\
\ion{Fe}{1}  & 4924.77 & 2.28 & $-$2.11 &   8.06 &    4.20 &  1 & 0.01 \\
\ion{Fe}{1}  & 4938.81 & 2.88 & $-$1.08 &  11.90 &    4.06 &  1 & 0.09 \\
\ion{Fe}{1}  & 4939.69 & 0.86 & $-$3.25 &  25.51 &    4.25 &  1 & 0.09 \\
\ion{Fe}{1}  & 4994.13 & 0.92 & $-$2.97 &  35.03 &    4.24 &  1 & 0.21 \\
\ion{Fe}{1}  & 5006.12 & 2.83 & $-$0.62 &  26.63 &    3.98 &  1 & 0.21 \\
\ion{Fe}{1}  & 5049.82 & 2.28 & $-$1.36 &  27.28 &    4.08 &  1 & 0.22 \\
\ion{Fe}{1}  & 5051.63 & 0.92 & $-$2.76 &  50.08 &    4.32 &  1 & 0.21 \\
\ion{Fe}{1}  & 5068.77 & 2.94 & $-$1.04 &  13.13 &    4.13 &  1 & 0.01 \\
\ion{Fe}{1}  & 5079.22 & 2.20 & $-$2.10 &  11.89 &    4.27 &  1 & 0.21 \\
\ion{Fe}{1}  & 5079.74 & 0.99 & $-$3.24 &  23.16 &    4.32 &  1 & 0.21 \\
\ion{Fe}{1}  & 5083.34 & 0.96 & $-$2.84 &  34.96 &    4.15 &  1 & 0.21 \\
\ion{Fe}{1}  & 5127.36 & 0.92 & $-$3.25 &  21.93 &    4.21 &  1 & 0.21 \\
\ion{Fe}{1}  & 5133.69 & 4.18 &    0.36 &  14.33 &    4.21 &  1 & 0.24 \\
\ion{Fe}{1}  & 5150.84 & 0.99 & $-$3.04 &  25.84 &    4.18 &  1 & \nodata \\
\ion{Fe}{1}  & 5151.91 & 1.01 & $-$3.32 &  15.19 &    4.19 &  1 & 0.21 \\
\ion{Fe}{1}  & 5171.60 & 1.49 & $-$1.72 &  62.58 &    4.21 &  1 & 0.21 \\
\ion{Fe}{1}  & 5191.45 & 3.04 & $-$0.55 &  33.40 &    4.29 &  1 & 0.21 \\
\ion{Fe}{1}  & 5192.34 & 3.00 & $-$0.42 &  23.64 &    3.89 &  1 & 0.21 \\
\ion{Fe}{1}  & 5194.94 & 1.56 & $-$2.02 &  40.55 &    4.15 &  1 & 0.22 \\
\ion{Fe}{1}  & 5198.71 & 2.22 & $-$2.09 &   7.08 &    4.03 &  1 & 0.21 \\
\ion{Fe}{1}  & 5202.34 & 2.18 & $-$1.87 &  17.98 &    4.22 &  1 & 0.22 \\
\ion{Fe}{1}  & 5216.27 & 1.61 & $-$2.08 &  30.13 &    4.05 &  1 & 0.21 \\
\ion{Fe}{1}  & 5232.94 & 2.94 & $-$0.06 &  46.07 &    3.93 &  1 & 0.21 \\
\ion{Fe}{1}  & 5263.31 & 3.27 & $-$0.87 &   9.29 &    4.17 &  1 & 0.22 \\
\ion{Fe}{1}  & 5266.56 & 3.00 & $-$0.38 &  24.30 &    3.87 &  1 & 0.21 \\
\ion{Fe}{1}  & 5283.62 & 3.24 & $-$0.45 &  15.66 &    3.97 &  1 & 0.22 \\
\ion{Fe}{1}  & 5324.18 & 3.21 & $-$0.11 &  25.73 &    3.87 &  1 & 0.23 \\
\ion{Fe}{1}  & 5339.93 & 3.27 & $-$0.63 &  11.80 &    4.04 &  1 & 0.22 \\
\ion{Fe}{1}  & 5383.37 & 4.31 &    0.64 &  11.28 &    3.95 &  1 & 0.26 \\
\ion{Fe}{1}  & 5415.20 & 4.39 &    0.64 &  13.86 &    4.15 &  1 & 0.27 \\
\ion{Fe}{1}  & 5497.52 & 1.01 & $-$2.82 &  38.38 &    4.22 &  1 & 0.00 \\
\ion{Fe}{1}  & 5501.47 & 0.96 & $-$3.05 &  34.53 &    4.32 &  1 & 0.00 \\
\ion{Fe}{1}  & 5506.78 & 0.99 & $-$2.79 &  45.96 &    4.31 &  1 & 0.01 \\
\ion{Fe}{1}  & 5586.76 & 3.37 & $-$0.11 &  21.98 &    3.95 &  1 & 0.23 \\
\ion{Fe}{1}  & 6136.61 & 2.45 & $-$1.41 &  23.98 &    4.19 &  1 & 0.20 \\
\ion{Fe}{1}  & 6137.69 & 2.59 & $-$1.35 &  12.49 &    3.95 &  1 & 0.19 \\
\ion{Fe}{1}  & 6191.56 & 2.43 & $-$1.42 &  19.68 &    4.06 &  1 & 0.22 \\
\ion{Fe}{1}  & 6230.72 & 2.56 & $-$1.28 &  22.45 &    4.15 &  1 & 0.22 \\
\ion{Fe}{1}  & 6252.56 & 2.40 & $-$1.77 &  12.82 &    4.15 &  1 & 0.22 \\
\ion{Fe}{1}  & 6265.13 & 2.18 & $-$2.54 &   5.74 &    4.27 &  1 & 0.22 \\
\ion{Fe}{1}  & 6393.60 & 2.43 & $-$1.58 &  15.87 &    4.09 &  1 & 0.23 \\
\ion{Fe}{1}  & 6411.65 & 3.65 & $-$0.59 &   9.74 &    4.30 &  1 & 0.24 \\
\ion{Fe}{1}  & 6421.35 & 2.28 & $-$2.01 &  11.67 &    4.19 &  1 & 0.00 \\
\ion{Fe}{1}  & 6430.85 & 2.18 & $-$1.95 &  16.05 &    4.17 &  1 & 0.00 \\
\ion{Fe}{1}  & 6494.98 & 2.40 & $-$1.24 &  33.30 &    4.14 &  1 & 0.23 \\
\ion{Fe}{1}  & 6592.91 & 2.73 & $-$1.47 &  10.70 &    4.13 &  1 & 0.17 \\
\ion{Fe}{1}  & 6677.99 & 2.69 & $-$1.42 &  17.55 &    4.28 &  1 & 0.22 \\
\ion{Fe}{1}  & 7511.02 & 4.18 &    0.12 &   9.09 &    4.13 &  1 & \nodata \\
\ion{Fe}{2}  & 4233.16 & 2.58 & $-$2.02 &  46.12 &    4.19 &  1 & \nodata \\
\ion{Fe}{2}  & 4303.17 & 2.70 & $-$2.52 &  16.63 &    4.13 &  1 & \nodata \\
\ion{Fe}{2}  & 4385.38 & 2.78 & $-$2.64 &  13.52 &    4.23 &  1 & \nodata \\
\ion{Fe}{2}  & 4416.82 & 2.78 & $-$2.57 &  15.46 &    4.23 &  1 & \nodata \\
\ion{Fe}{2}  & 4508.28 & 2.86 & $-$2.42 &  15.09 &    4.15 &  1 & \nodata \\
\ion{Fe}{2}  & 4515.34 & 2.84 & $-$2.60 &  10.87 &    4.14 &  1 & \nodata \\
\ion{Fe}{2}  & 4555.89 & 2.83 & $-$2.40 &  16.57 &    4.15 &  1 & \nodata \\
\ion{Fe}{2}  & 4583.83 & 2.81 & $-$1.94 &  40.30 &    4.22 &  1 & \nodata \\
\ion{Fe}{2}  & 5197.58 & 3.23 & $-$2.22 &  11.79 &    4.22 &  1 & \nodata \\
\ion{Fe}{2}  & 5234.63 & 3.22 & $-$2.18 &  12.35 &    4.19 &  1 & \nodata \\
\ion{Fe}{2}  & 5276.00 & 3.20 & $-$2.01 &  19.03 &    4.22 &  1 & \nodata \\
\ion{Co}{1}  & 3845.47 & 0.92 &    0.06 &  55.89 &    1.65 &  1 & 0.83 \\
\ion{Co}{1}  & 3894.08 & 1.05 &    0.12 &  51.22 &    1.63 &  1 & 0.82 \\
\ion{Co}{1}  & 3995.31 & 0.92 & $-$0.18 &  51.27 &    1.76 &  1 & 0.70 \\
\ion{Ni}{1}  & 5476.90 & 1.83 & $-$0.78 &  31.49 &    2.74 &  1 & \nodata \\
\ion{Zn}{1}  & 4722.15 & 4.03 & $-$0.37 &   syn  &    1.36 &  1 & \nodata \\
\ion{Sr}{2}  & 4077.71 & 0.00 &    0.15 &   syn  &    0.37 &  2 & \nodata \\
\ion{Sr}{2}  & 4215.52 & 0.00 & $-$0.17 &   syn  &    0.37 &  2 & \nodata \\
\ion{Y}{2}   & 3747.55 & 0.10 & $-$0.95 &   syn  & $-$0.69 &  3 & \nodata \\
\ion{Y}{2}   & 4398.01 & 0.13 & $-$0.75 &   syn  & $-$0.67 &  3 & \nodata \\
\ion{Y}{2}   & 4883.68 & 1.08 &    0.19 &   syn  & $-$0.79 &  3 & \nodata \\
\ion{Y}{2}   & 4900.12 & 1.03 &    0.03 &   syn  & $-$0.76 &  3 & \nodata \\
\ion{Y}{2}   & 5205.72 & 1.03 & $-$0.28 &   syn  & $-$0.76 &  3 & \nodata \\
\ion{Y}{2}   & 5662.92 & 1.94 &    0.34 &   syn  & $-$0.76 &  3 & \nodata \\
\ion{Zr}{2}  & 3836.76 & 0.56 & $-$0.12 &   syn  & $-$0.27 &  4 & \nodata \\
\ion{Zr}{2}  & 4149.20 & 0.80 & $-$0.04 &   syn  & $-$0.17 &  4 & \nodata \\
\ion{Zr}{2}  & 4161.20 & 0.71 & $-$0.59 &   syn  & $-$0.17 &  4 & \nodata \\
\ion{Zr}{2}  & 4208.98 & 0.71 & $-$0.51 &   syn  & $-$0.17 &  4 & \nodata \\
\ion{Ba}{2}  & 5853.68 & 0.60 & $-$0.91 &   syn  & $-$0.10 &  5 & \nodata \\
\ion{Ba}{2}  & 6141.71 & 0.70 & $-$0.01 &   syn  &    0.08 &  1 & \nodata \\
\ion{Ba}{2}  & 6496.90 & 0.60 & $-$0.37 &   syn  &    0.13 &  1 & \nodata \\
\ion{La}{2}  & 3988.51 & 0.40 &    0.21 &   syn  & $-$1.05 &  6 & \nodata \\
\ion{La}{2}  & 3995.74 & 0.17 & $-$0.06 &   syn  & $-$0.95 &  6 & \nodata \\
\ion{La}{2}  & 4077.34 & 0.24 & $-$0.06 &   syn  & $-$1.05 &  6 & \nodata \\
\ion{La}{2}  & 4086.71 & 0.00 & $-$0.07 &   syn  & $-$1.02 &  6 & \nodata \\
\ion{La}{2}  & 4123.22 & 0.32 &    0.13 &   syn  & $-$1.12 &  6 & \nodata \\
\ion{La}{2}  & 4429.91 & 0.24 & $-$0.35 &   syn  & $-$1.00 &  6 & \nodata \\
\ion{La}{2}  & 4920.98 & 0.13 & $-$0.58 &   syn  & $-$0.98 &  6 & \nodata \\
\ion{La}{2}  & 4921.78 & 0.24 & $-$0.45 &   syn  & $-$0.90 &  6 & \nodata \\
\ion{Ce}{2}  & 3942.74 & 0.86 &    0.69 &   syn  & $-$0.52 &  7 & \nodata \\
\ion{Ce}{2}  & 4073.47 & 0.48 &    0.21 &   syn  & $-$0.67 &  7 & \nodata \\
\ion{Ce}{2}  & 4075.70 & 0.70 &    0.23 &   syn  & $-$0.57 &  7 & \nodata \\
\ion{Ce}{2}  & 4137.64 & 0.52 &    0.40 &   syn  & $-$0.62 &  7 & \nodata \\
\ion{Ce}{2}  & 4165.60 & 0.91 &    0.52 &   syn  & $-$0.67 &  7 & \nodata \\
\ion{Ce}{2}  & 4222.60 & 0.12 & $-$0.15 &   syn  & $-$0.67 &  7 & \nodata \\
\ion{Ce}{2}  & 4449.33 & 0.61 &    0.04 &   syn  & $-$0.72 &  7 & \nodata \\
\ion{Ce}{2}  & 4450.73 & 0.68 & $-$0.17 &   syn  & $-$0.59 &  1 & \nodata \\
\ion{Ce}{2}  & 4562.36 & 0.48 &    0.21 &   syn  & $-$0.69 &  7 & \nodata \\
\ion{Ce}{2}  & 4628.16 & 0.52 &    0.14 &   syn  & $-$0.62 &  7 & \nodata \\
\ion{Pr}{2}  & 4222.95 & 0.06 &    0.23 &   syn  & $-$1.13 &  8 & \nodata \\
\ion{Pr}{2}  & 4225.32 & 0.00 &    0.32 &   syn  & $-$1.13 &  8 & \nodata \\
\ion{Pr}{2}  & 4408.81 & 0.00 &    0.05 &   syn  & $-$1.18 &  8 & \nodata \\
\ion{Nd}{2}  & 3826.41 & 0.06 & $-$0.41 &   syn  & $-$0.56 &  9 & \nodata \\
\ion{Nd}{2}  & 3838.98 & 0.00 & $-$0.24 &   syn  & $-$0.63 &  9 & \nodata \\
\ion{Nd}{2}  & 3879.54 & 0.32 & $-$0.21 &   syn  & $-$0.58 &  9 & \nodata \\
\ion{Nd}{2}  & 3880.37 & 0.18 & $-$0.59 &   syn  & $-$0.53 &  1 & \nodata \\
\ion{Nd}{2}  & 3880.77 & 0.06 & $-$0.31 &   syn  & $-$0.58 &  1 & \nodata \\
\ion{Nd}{2}  & 3900.22 & 0.47 &    0.10 &   syn  & $-$0.63 &  9 & \nodata \\
\ion{Nd}{2}  & 3990.10 & 0.47 &    0.13 &   syn  & $-$0.63 &  9 & \nodata \\
\ion{Nd}{2}  & 3991.74 & 0.00 & $-$0.26 &   syn  & $-$0.58 &  9 & \nodata \\
\ion{Nd}{2}  & 4004.00 & 0.06 & $-$0.57 &   syn  & $-$0.63 &  9 & \nodata \\
\ion{Nd}{2}  & 4012.24 & 0.63 &    0.81 &   syn  & $-$0.61 &  9 & \nodata \\
\ion{Nd}{2}  & 4051.14 & 0.38 & $-$0.30 &   syn  & $-$0.53 &  9 & \nodata \\
\ion{Nd}{2}  & 4059.95 & 0.20 & $-$0.52 &   syn  & $-$0.55 &  9 & \nodata \\
\ion{Nd}{2}  & 4061.08 & 0.47 &    0.55 &   syn  & $-$0.65 &  1 & \nodata \\
\ion{Nd}{2}  & 4109.45 & 0.32 &    0.35 &   syn  & $-$0.53 &  9 & \nodata \\
\ion{Nd}{2}  & 4110.47 & 0.00 & $-$0.71 &   syn  & $-$0.58 &  1 & \nodata \\
\ion{Nd}{2}  & 4303.57 & 0.00 &    0.08 &   syn  & $-$0.48 &  9 & \nodata \\
\ion{Nd}{2}  & 4451.98 & 0.00 & $-$1.10 &   syn  & $-$0.58 &  9 & \nodata \\
\ion{Nd}{2}  & 4465.06 & 0.00 & $-$1.36 &   syn  & $-$0.58 &  9 & \nodata \\
\ion{Nd}{2}  & 4465.59 & 0.18 & $-$1.10 &   syn  & $-$0.63 &  9 & \nodata \\
\ion{Nd}{2}  & 4706.54 & 0.00 & $-$0.71 &   syn  & $-$0.51 &  9 & \nodata \\
\ion{Nd}{2}  & 4825.48 & 0.18 & $-$0.42 &   syn  & $-$0.58 &  9 & \nodata \\
\ion{Nd}{2}  & 5249.58 & 0.98 &    0.20 &   syn  & $-$0.63 &  9 & \nodata \\
\ion{Sm}{2}  & 3922.39 & 0.38 & $-$0.26 &   syn  & $-$1.04 & 10 & \nodata \\
\ion{Sm}{2}  & 4188.13 & 0.54 & $-$0.44 &   syn  & $-$0.84 & 10 & \nodata \\
\ion{Sm}{2}  & 4279.74 & 0.18 & $-$1.24 &   syn  & $-$0.89 &  1 & \nodata \\
\ion{Sm}{2}  & 4280.78 & 0.48 &    0.06 &   syn  & $-$0.89 & 10 & \nodata \\
\ion{Sm}{2}  & 4318.94 & 0.28 & $-$0.25 &   syn  & $-$0.89 & 10 & \nodata \\
\ion{Sm}{2}  & 4433.89 & 0.43 & $-$0.19 &   syn  & $-$0.94 & 10 & \nodata \\
\ion{Sm}{2}  & 4434.32 & 0.38 & $-$0.07 &   syn  & $-$0.94 & 10 & \nodata \\
\ion{Sm}{2}  & 4467.34 & 0.66 &    0.15 &   syn  & $-$1.14 & 10 & \nodata \\
\ion{Sm}{2}  & 4472.41 & 0.18 & $-$0.96 &   syn  & $-$0.89 & 10 & \nodata \\
\ion{Sm}{2}  & 4519.63 & 0.54 & $-$0.35 &   syn  & $-$0.89 & 10 & \nodata \\
\ion{Sm}{2}  & 4523.91 & 0.43 & $-$0.39 &   syn  & $-$1.04 & 10 & \nodata \\
\ion{Sm}{2}  & 4566.20 & 0.33 & $-$0.59 &   syn  & $-$0.92 & 10 & \nodata \\
\ion{Sm}{2}  & 4577.69 & 0.25 & $-$0.65 &   syn  & $-$0.94 & 10 & \nodata \\
\ion{Sm}{2}  & 4615.44 & 0.54 & $-$0.69 &   syn  & $-$0.87 & 10 & \nodata \\
\ion{Sm}{2}  & 4669.39 & 0.10 & $-$0.60 &   syn  & $-$0.89 & 10 & \nodata \\
\ion{Sm}{2}  & 4669.64 & 0.28 & $-$0.53 &   syn  & $-$0.84 & 10 & \nodata \\
\ion{Sm}{2}  & 4674.59 & 0.18 & $-$0.56 &   syn  & $-$0.87 & 10 & \nodata \\
\ion{Eu}{2}  & 3724.93 & 0.00 & $-$0.09 &   syn  & $-$1.17 & 11 & \nodata \\
\ion{Eu}{2}  & 3819.67 & 0.00 &    0.51 &   syn  & $-$1.27 & 11 & \nodata \\
\ion{Eu}{2}  & 3907.11 & 0.21 &    0.17 &   syn  & $-$1.36 & 11 & \nodata \\
\ion{Eu}{2}  & 3971.97 & 0.21 &    0.27 &   syn  & $-$1.27 & 11 & \nodata \\
\ion{Eu}{2}  & 4129.72 & 0.00 &    0.22 &   syn  & $-$1.28 & 11 & \nodata \\
\ion{Eu}{2}  & 4205.04 & 0.00 &    0.21 &   syn  & $-$1.26 & 11 & \nodata \\
\ion{Eu}{2}  & 4435.58 & 0.21 & $-$0.11 &   syn  & $-$1.20 & 11 & \nodata \\
\ion{Eu}{2}  & 6645.06 & 1.38 &    0.12 &   syn  & $-$1.23 & 11 & \nodata \\
\ion{Gd}{2}  & 3844.58 & 0.14 & $-$0.46 &   syn  & $-$0.76 &  1 & \nodata \\
\ion{Gd}{2}  & 3957.67 & 0.60 & $-$0.25 &   syn  & $-$0.68 & 12 & \nodata \\
\ion{Gd}{2}  & 4049.42 & 0.66 & $-$0.08 &   syn  & $-$0.72 & 12 & \nodata \\
\ion{Gd}{2}  & 4049.85 & 0.99 &    0.48 &   syn  & $-$0.66 & 12 & \nodata \\
\ion{Gd}{2}  & 4085.56 & 0.73 & $-$0.01 &   syn  & $-$0.78 & 12 & \nodata \\
\ion{Gd}{2}  & 4130.37 & 0.73 &    0.14 &   syn  & $-$0.73 &  1 & \nodata \\
\ion{Gd}{2}  & 4132.26 & 0.60 & $-$0.15 &   syn  & $-$0.75 & 12 & \nodata \\
\ion{Gd}{2}  & 4191.07 & 0.43 & $-$0.48 &   syn  & $-$0.78 & 12 & \nodata \\
\ion{Gd}{2}  & 4251.73 & 0.38 & $-$0.22 &   syn  & $-$0.78 & 12 & \nodata \\
\ion{Tb}{2}  & 3874.17 & 0.00 &    0.27 &   syn  & $-$1.35 & 13 & \nodata \\
\ion{Dy}{2}  & 3944.68 & 0.00 &    0.11 &   syn  & $-$0.30 & 14 & \nodata \\
\ion{Dy}{2}  & 3978.56 & 0.92 &    0.22 &   syn  & $-$0.50 & 14 & \nodata \\
\ion{Dy}{2}  & 3983.65 & 0.54 & $-$0.31 &   syn  & $-$0.50 & 14 & \nodata \\
\ion{Dy}{2}  & 3996.69 & 0.59 & $-$0.26 &   syn  & $-$0.60 & 14 & \nodata \\
\ion{Dy}{2}  & 4050.57 & 0.59 & $-$0.47 &   syn  & $-$0.65 & 14 & \nodata \\
\ion{Dy}{2}  & 4073.12 & 0.54 & $-$0.32 &   syn  & $-$0.55 & 14 & \nodata \\
\ion{Dy}{2}  & 4077.97 & 0.10 & $-$0.04 &   syn  & $-$0.40 & 14 & \nodata \\
\ion{Dy}{2}  & 4103.31 & 0.10 & $-$0.38 &   syn  & $-$0.30 & 14 & \nodata \\
\ion{Dy}{2}  & 4449.70 & 0.00 & $-$1.03 &   syn  & $-$0.40 & 14 & \nodata \\
\ion{Ho}{2}  & 3810.71 & 0.00 &    0.19 &   syn  & $-$1.27 & 15 & \nodata \\
\ion{Ho}{2}  & 3890.97 & 0.08 &    0.46 &   syn  & $-$1.44 & 15 & \nodata \\
\ion{Ho}{2}  & 4045.45 & 0.00 & $-$0.05 &   syn  & $-$1.27 & 15 & \nodata \\
\ion{Er}{2}  & 3830.48 & 0.00 & $-$0.22 &   syn  & $-$0.85 & 16 & \nodata \\
\ion{Er}{2}  & 3880.61 & 0.64 & $-$0.25 &   syn  & $-$0.71 & 16 & \nodata \\
\ion{Er}{2}  & 3896.23 & 0.06 & $-$0.12 &   syn  & $-$0.85 & 16 & \nodata \\
\ion{Er}{2}  & 3938.63 & 0.00 & $-$0.52 &   syn  & $-$0.78 &  1 & \nodata \\
\ion{Tm}{2}  & 3795.76 & 0.03 & $-$0.23 &   syn  & $-$1.55 & 17 & \nodata \\
\ion{Tm}{2}  & 3848.02 & 0.00 & $-$0.14 &   syn  & $-$1.75 & 17 & \nodata \\
\ion{Tm}{2}  & 3996.51 & 0.00 & $-$1.20 &   syn  & $-$1.68 & 17 & \nodata \\
\ion{Tm}{2}  & 4242.15 & 0.03 & $-$0.95 &   syn  & $-$1.60 & 17 & \nodata \\
\ion{Yb}{2}  & 3694.19 & 0.00 & $-$0.30 &   syn  & $-$1.06 & 18 & \nodata \\
\ion{Hf}{2}  & 4093.15 & 0.45 & $-$1.15 &   syn  & $-$1.15 & 19 & \nodata \\
\ion{Os}{1}  & 4260.80 & 0.00 & $-$1.43 &   syn  & $-$0.25 & 20 & \nodata \\
\ion{Os}{1}  & 4420.47 & 0.00 & $-$1.20 &   syn  & $-$0.18 & 20 & \nodata \\
\ion{Ir}{1}  & 3800.12 & 0.00 & $-$1.43 &   syn  & $-$0.35 & 21 & \nodata \\
\ion{Th}{2}  & 4019.13 & 0.00 & $-$0.23 &   syn  & $-$1.13 & 22 & \nodata \\
\ion{Th}{2}  & 4086.52 & 0.00 & $-$0.93 &   syn  & $-$1.23 & 22 & \nodata \\
\ion{Th}{2}  & 4094.75 & 0.00 & $-$0.88 &   syn  & $-$1.28 & 22 & \nodata \\
\enddata
\tablerefs{
 1: \citet{placco2021,placco2021a};
 2: \citet{nist};
 3: \citet{biemont2011};
 4: \citet{ljung2006};
 5: \citet{nist}, using HFS/IS from \citet{mcwilliam1998} when available;
 6: \citet{lawler2001}, using HFS from \citet{ivans2006} when available;
 7: \citet{lawler2009};
 8: \citet{li2007}, using HFS from \citet{sneden2009};
 9: \citet{denhartog2003}, using HFS/IS from \citet{roederer2008} when available;
10: \citet{lawler2006}, using HFS/IS from \citet{roederer2008} when available;
11: \citet{lawler2001}, using HFS/IS from \citet{ivans2006};
12: \citet{denhartog2006};
13: \citet{lawler2001b}, using HFS from \citet{lawler2001b};
14: \citet{wickliffe2000};
15: \citet{lawler2004}, using HFS from \citet{sneden2009};
16: \citet{lawler2008};
17: \citet{wickliffe1997}, using HFS from \citet{sneden2009};
18: \citet{sneden2009} for log(gf) value and HFS/IS;
19: \citet{lawler2007};
20: \citet{quinet2006};
21: \citet{xu2007}, using HFS/IS from \citet{cowan2005};
22: \citet{nilsson2002}.
NLTE corrections -- 
\ion{Na}{1}: \citet{lind2011};
\ion{Mg}{1}: \citet{bergemann2015};
\ion{Al}{1}: \citet{nordlander2017b};
\ion{Si}{1}: \citet{bergemann2013};
\ion{Ca}{1}: \citet{mashonkina2007};
\ion{Ti}{1}: \citet{bergemann2011};
\ion{Ti}{2}: \citet{bergemann2011};
\ion{Cr}{1}: \citet{bergemann2010};
\ion{Mn}{1}: \citet{bergemann2019};
\ion{Fe}{1}: \citet{bergemann2012b};
\ion{Co}{1}: \citet{bergemann2010b}
}
\end{deluxetable}

\end{document}